\documentclass[twocolumn,floatfix,prl,aps,showpacs, longbibliography, superscriptaddress]{revtex4-2}
\usepackage{graphicx,amsmath,amssymb, color}
\usepackage{nicefrac}
\usepackage[titletoc,title]{appendix}
\usepackage[colorlinks,bookmarks=true,citecolor=blue,linkcolor=red,urlcolor=blue]{hyperref}
\usepackage{orcidlink}
\usepackage{comment}

\begin{document}

\title{Fractional quantum Hall effect of partons and the nature of the 8/17 state in the zeroth Landau level of bilayer graphene}

\author{Ajit C. Balram\orcidlink{0000-0002-8087-6015}}
\email{cb.ajit@gmail.com}
\affiliation{Institute of Mathematical Sciences, CIT Campus, Chennai 600113, India}
\affiliation{Homi Bhabha National Institute, Training School Complex, Anushaktinagar, Mumbai 400094, India}

\author{Nicolas Regnault}
\email{nicolas.regnault@phys.ens.fr}
\affiliation{Laboratoire de Physique de l’Ecole normale sup\'erieure,
ENS, Universit\'e PSL, CNRS, Sorbonne Universit\'e,
Universit\'e Paris-Diderot, Sorbonne Paris Cit\'e, 75005 Paris, France}
\affiliation{Department of Physics, Princeton University, Princeton, New Jersey 08544, USA}

\date{\today}

\begin{abstract} 
We consider the fractional quantum Hall effect (FQHE) at the filling factor $8/17$, where signatures of incompressibility have been observed in the zeroth Landau level of bilayer graphene. We propose an Abelian state described by the ``$\overline{(8/3)}\bar{2}1^{3}$" parton wave function, where a parton itself forms an FQHE state. This state is topologically distinct from the $8/17$ Levin-Halperin state, a daughter state of the Moore-Read state. We carry out extensive numerical exact diagonalization of the Coulomb interaction at 8/17 in the zeroth Landau level of bilayer graphene but find that our results cannot conclusively determine the topological order of the underlying ground state. We work out the low-energy effective theory of the $\overline{(8/3)}\bar{2}1^{3}$ edge and make predictions for experimentally measurable properties of the state which can tell it apart from the 8/17 Levin-Halperin state. 
\end{abstract}

\maketitle

The fractional quantum Hall effect (FQHE)~\cite{Tsui82, Laughlin83} has underpinned major developments in the field of strongly interacting topological phases of matter. The FQHE is observed in cryogenic two-dimensional systems placed in a perpendicular magnetic field and manifests itself as precisely quantized plateaux in the Hall resistance at certain special fractions. The FQHE phenomena arise from complex many-particle correlations that are characterized by topological order and long-range quantum entanglement~\cite{Wen95}. Starting with the work of Laughlin~\cite{Laughlin83}, numerous schemes have been proposed to build these many-body correlations to capture the myriad of fractions observed. These include the Haldane-Halperin hierarchical construction~\cite{Haldane83, Halperin84}, Jain's composite fermion (CF)~\cite{Jain89} and parton~\cite{Jain89b} theories, deploying techniques of conformal field theory (CFT)~\cite{Hansson17}, and many more. 

The Haldane-Halperin hierarchy postulates that the quasiholes or quasiparticle excitations of a parent FQHE state condense into a daughter FQHE state and this scheme can produce candidate states for any odd-denominator fraction~\cite{Haldane83, Halperin84}. Jain's CF theory builds many-particle correlations by the process of vortex attachment wherein electrons bind zeros of the many-body wave function to turn into CFs which to the zeroth-order approximation are taken to be non-interacting~\cite{Jain89}. A vast majority of the FQHE phenomenology, especially that observed in the lowest Landau level (LLL), is well-described by the CF theory. In particular, the CF theory explains why the strongest fractions observed in the LLL take the form $\nu{=}n/(2pn{\pm}1)$, with $n,p$ positive integers, as these corresponds to the $\nu^{*}{=}n$ integer quantum Hall effect (IQH) states of CFs. The theory of free CFs also produces states only at odd-denominator fractions. 

The experimental observation of a plateau at filling factor $\nu{=}5/2$ in the second Landau level (SLL)~\cite{Willett87} showed that FQHE could also arise at even-denominator fractions. Moore and Read~\cite{Moore91}, using the methods of CFT, postulated a non-Abelian Pfaffian (Pf) wave function to describe this even-denominator state which was subsequently shown to give a good representation of the exact SLL Coulomb ground state at half-filling~\cite{Morf02, Balram20b}. Read and Rezayi~\cite{Read99} then generalized the Moore-Read construction to produce a family of states, some of which can capture other experimentally observed plateaux in the SLL~\cite{Rezayi09, Peterson15}. Some experimentally observed fractions in the SLL still lie beyond the purview of the aforementioned theories~\cite{Xia04, Kumar10}. In particular, signatures of FQHE have been observed in the zeroth LL (ZLL) of bilayer graphene (BLG), which is believed to stabilize states analogous to those in the SLL, at fractions $6/13$ and $8/17$ and their hole-conjugates $7/13$ and $9/17$~\cite{Zibrov16, Huang21, Assouline23, Hu23, Kumar24} (Throughout this work, states related by particle-hole symmetry would be considered on an equal footing and thus it suffices to just consider $\nu{\leq} 1/2$.). 

Levin and Halperin (LH)~\cite{Levin09a} carried out a hierarchical construction in which the quasiholes (qhs) or quasiparticles (qps) of the Pf state or its hole-conjugate called the anti-Pfaffian (aPf)~\cite{Levin07, Lee07}, condense into a daughter Laughlin state. At the first level of the hierarchy, condensing the qps and qhs of the Pf and aPf states produces four states that precisely occur at the above-mentioned fillings of $7/13$ (qps of Pf), $8/17$ (qhs of Pf), $6/13$ (qps of aPf), and $9/17$ (qhs of aPf). In this work, we focus on the $8/17$ state and test if the LH construction gives a viable candidate to describe the Coulomb ground state in the ZLL of BLG. We note that FQHE at $3/8$ in the SLL has been observed~\cite{Xia04, Kumar10} and this fraction still lies beyond the realm of all the theories mentioned above. 

Very recently, it has been observed that in wide quantum wells (WQWs) with increasing density, the \emph{LLL} states at $\nu{=}7/13$ and $\nu{=}8/17$ become stronger in conjunction with the strengthening of the $1/2$ state~\cite{Singh23}. Theoretical calculations~\cite{Zhao21, Sharma23} and the experimental observations suggest that the 1/2 state lies in the topological phase of the Pf state. Therefore, the $7/13$ and $8/17$ states seen in the LLL in WQWs are expected to be described by the LH daughter states of the Pf. 

In recent work, a parton~\cite{Jain89b} based sequence has been proposed to capture \emph{all} of the experimentally observed fractions in the SLL in the vicinity of half-filling~\cite{Balram18, Balram19, Balram20a, Bose23} (The only other states observed in the SLL lie in the $n/(4n{\pm}1)$ Jain sequence and these are expected to be analogous to their LLL counterparts~\cite{Balram20a}.). Interestingly, $8/17$ is the only fraction where signatures of incompressibility have been seen that does not fit the proposed parton sequence. Other than the $8/17$ Jain state, which is not expected to be relevant for SLL, no parton sequence is known that has $8/17$ as its member. In the parton theory~\cite{Jain89b}, FQHE states are obtained from products of IQHE states that the non-interacting partons fill. We generalize this construction to postulate that in certain scenarios, the partons themselves can be strongly interacting and can undergo FQHE. Using this idea, we produce a new candidate state at $8/17$ that is closely related to the SLL parton sequence. We shall only consider single-component physics since the experiments are done in the ZLL of BLG with a fixed spin and valley index.

{\bf \em Parton theory.---}Jain generalized his CF construction to the parton theory in which FQHE states of electrons are constructed from IQHE states of sub-particles called partons~\cite{Jain89b}. The electron is partitioned into odd-$l$ species of partons, labeled by $\alpha{=}1,2,{\cdots},l$, and the $\alpha$ parton species is placed in an IQHE state at filling $n_{\alpha}$. This parton state, denoted by ``$n_{1} n_{2} {\cdots} n_{l}$", is described by the wave function~\cite{Jain89b}
\begin{equation}
	\Psi^{n_{1}n_{2}\cdots n_{l}}_{\nu} = \mathcal{P}_{\rm LLL} \prod_{\alpha=1}^{l} \Phi_{n_{\alpha}},
	\label{eq: parton_general}
\end{equation}
where $\Phi_{n}$ is the Slater determinant state of $n$ filled LLs of electrons with $\Phi_{{-}n}{\equiv} \Phi_{\bar{n}} {=}[\Phi_{n}]^{*}$, and $\mathcal{P}_{\rm LLL}$ implements projection to the LLL as is required in the infinite field limit. Using the fact that the partons are exposed to the same external magnetic field as the electrons and have the same density as the electrons, one can show that the charge of the $\alpha$ parton species $q_{\alpha}{=}{-}e \nu/n_{\alpha}$, where $\nu{=}[\sum_{\alpha{=}1}^{l}n^{-1}_{\alpha}]^{-1}$~\cite{Jain89b}. The Wen-Zee shift~\cite{Wen92} of the parton state of Eq.~\eqref{eq: parton_general} is $\mathcal{S}{=}\sum_{\alpha{=}1}^{l}n_{\alpha}$ and is thus always an integer. However, an FQHE state can also have a fractional shift~\cite{Levin09a, Mukherjee14, Balram15a, Balram16c}, and such states (in particular, a family of LH states that include $8/17$ have fractional shifts) are not captured by the wave function of Eq.~\eqref{eq: parton_general}.

Many well-known FQHE states can be obtained as parton states. The $1/3$ Laughlin state~\cite{Laughlin83} is a $111$ parton state described by the wave function $\Psi^{\rm Laughlin}_{1/3}{=}\Phi^{3}_{1}$. The $n/(2pn{\pm}1)$ Jain states~\cite{Jain89} are ${\pm n}11{\cdots}$ parton states and these are described by the wave function $\Psi^{\rm Jain}_{n/(2pn{\pm}1)}{=}\mathcal{P}_{\rm LLL}\Phi_{{\pm} n}\Phi^{2p}_{1}$. Recently, many parton states that lie beyond those captured by free CFs, have been constructed to describe the ground state and also the excitations of many FQHE states~\cite{Wu17, Kim19, Faugno19, Faugno20a, Balram20, Balram21a, Balram21b, Balram21c, Balram21d, Sharma22, Bose23}.

Many states observed in the SLL can be described by the $\bar{n}\bar{2}1^{3}$ states that are described by the wave function
\begin{equation}
	\Psi^{\bar{n}\bar{2}1^{3}}_{\nu=2n/(5n-2)} = \mathcal{P}_{\rm LLL} \left[\Phi_{n}\right]^{*}[\Phi_{2}]^{*}\Phi^{3}_{1} \sim \frac{\Psi^{\rm Jain}_{n/(2n-1)}\Psi^{\rm Jain}_{2/3}}{\Phi_{1}},
	\label{eq: parton_barnbar2111}
\end{equation}
where $\Psi^{\rm Jain}_{n/(2n{-}1)}$ is the $n/(2n{-}1)$ Jain state~\cite{Jain89}. The $\sim$ sign in Eq.~\eqref{eq: parton_8_17_bar_8_3_bar2111} indicates that the states on the two sides of the sign differ in the details of how the LLL projection is carried out. Although such details do affect the microscopic form of the wave function, we anticipate that the universality class of the underlying topological phase remains unchanged~\cite{Balram16b, Anand22}. Only the wave function given in the right-most side of Eq.~\eqref{eq: parton_barnbar2111} is readily amenable to numerics and, thus, it is this form that we shall use in all our numerical calculations.

The $\bar{1}\bar{2}1^{3}$ is topologically equivalent (and nearly identical) to the $2/3$ Jain state~\cite{Balram16b} and the latter is known to give a reasonable description of the Coulomb ground state in the entire ZLL of BLG~\cite{Balram21b}. The $\bar{2}^{2}1^{3}$ state occurs at half-filling and lies in the same universality class as the aPf~\cite{Balram18}. Encouragingly, the $\bar{2}^{2}1^{3}$ provides a better representation of the Coulomb ground state in the ZLL of BLG near the SLL Coulomb point than the Pf state~\cite{Balram21b}. The $n{=}3$ member of the sequence given in Eq.~\eqref{eq: parton_barnbar2111} is a candidate state that occurs at $6/13$ where FQHE has been observed both in the SLL~\cite{Kumar10} and the ZLL of BLG~\cite{Zibrov16, Huang21, Assouline23, Hu23, Kumar24}. The $\bar{3}\bar{2}1^{3}$ wave function has been shown to give a good description of the exact SLL Coulomb ground state~\cite{Balram18a} and in the Supplemental Material (SM)~\cite{SM}, we show that it gives a good description of the ground state in the ZLL of BLG too. It turns out that the $\bar{3}\bar{2}1^{3}$ state lies in the same universality class as the corresponding LH state~\cite{Balram18a}.

Surprisingly, although the fraction $8/17$ does not appear for any integer $n$ in the sequence described in Eq.~\eqref{eq: parton_barnbar2111} it does occur if we set $n{=}8/3$. For the 8/17 FQHE state of our interest, we thus consider the parton state denoted as ``$\overline{(8/3)}\bar{2}1^{3}$" and described by the wave function
\begin{equation}
\Psi^{\overline{(8/3)}\bar{2}1^{3}}_{8/17} = \mathcal{P}_{\rm LLL} \left[\Phi_{8/3}\right]^{*}[\Phi_{2}]^{*}\Phi^{3}_{1} \sim \frac{\Psi_{8/13}\Psi^{\rm Jain}_{2/3}}{\Phi_{1}},
\label{eq: parton_8_17_bar_8_3_bar2111}
\end{equation}
where $\Psi_{8/13}{\equiv} \mathcal{P}_{\rm LLL}\Phi^{2}_{1}[\Phi_{8/3}]^{*}$ and the wave function $\Phi_{8/3}$ is constructed by filling the lowest two LLs of electrons and forming the $2/3$ Jain state (or equivalently, the particle-hole conjugate of the $1/3$ Laughlin state) in the third LL. The 8/3 wave function cannot be broken down into a product of IQHE states as it has a fractional shift of 5/2. The 8/17 state of Eq.~\eqref{eq: parton_8_17_bar_8_3_bar2111} occurs at shift ${-}3/2$ which is different from the LH state (shift 5/2) and therefore the two states are topologically distinct.

To draw parallels with CFs, we mention here that many FQHE states of electrons can be understood as arising from FQHE of CFs that stems from the residual interaction between them~\cite{Mukherjee14, Balram15, Balram16c}. These include fractions such as $4/11$, $5/13$ and $4/13$, where FQHE has been observed in the LLL~\cite{Pan03, Samkharadze15b, Pan15, Kumar18}. Recently, it has been proposed that these fractions can be understood as arising from IQHE of partons~\cite{Balram21a, Balram21c, Dora22} but the 8/17 state of our interest is outside the purview of IQHE of partons. 

{\bf \em Numerical results.---}All our calculations are carried out on the Haldane sphere~\cite{Haldane83} in which $N$ electrons reside on a spherical surface that is threaded by a flux of strength $2Qhc/e$ emanating from a magnetic monopole placed at the center of the sphere. The radius of the sphere $R$ is related to the flux as $R{=}\sqrt{Q}\ell$, where $\ell{=}\sqrt{\hbar c/(e B)}$ is the magnetic length at the field $B$. A quantum Hall state on the sphere occurs when $2Q{=}N/\nu{-}\mathcal{S}$, where $\mathcal{S}$ is the Wen-Zee shift~\cite{Wen92}, a topological quantum number that is a characteristic feature of the state. Owing to the rotational symmetry, the total orbital angular momentum $L$ and its $z$-component $L_{z}$ are good quantum numbers. In particular, incompressible quantum Hall liquid states are uniform i.e., have $L{=}0$. 

In the disk geometry, the electron-electron interaction in the ZLL of BLG is simulated by the form factor (the magnetic length is set to unity)
\begin{equation}
  F^{\mathcal{N}=1}(k) = \bigg[\sin^2(\theta)L_1\bigg(\frac{k^2}{2}\bigg) + \cos^2(\theta)L_0\bigg(\frac{k^2}{2}\bigg)\bigg]^2
  \label{eq: form_factor_N_1_BLG},
\end{equation}
where $k$ is the momentum, $L_m(x)$ is the $m$th ordered Laguerre polynomial, and $\theta$ is a parameter that depends on the external magnetic field. For typical BLG samples, $B{=}93.06 [\cot(\theta)]^{2}$ [in Tesla]~\cite{Jung14, Faugno21}. Using the form factor given in Eq.~\eqref{eq: form_factor_N_1_BLG}, one can compute the disk pseudopotentials in the ZLL of BLG~\cite{Faugno21} and an analogous computation can be carried out to obtain the spherical pseudopotentials~\cite{Balram21b}. The contribution of the positively charged background, which is required to estimate the charge gap shown below, is accounted for by computing the charging energy using these pseudopotentials~\cite{Balram20b}.

\begin{figure}[htpb]
	\begin{center}
		\includegraphics[width=0.49\textwidth,height=0.24\textwidth]{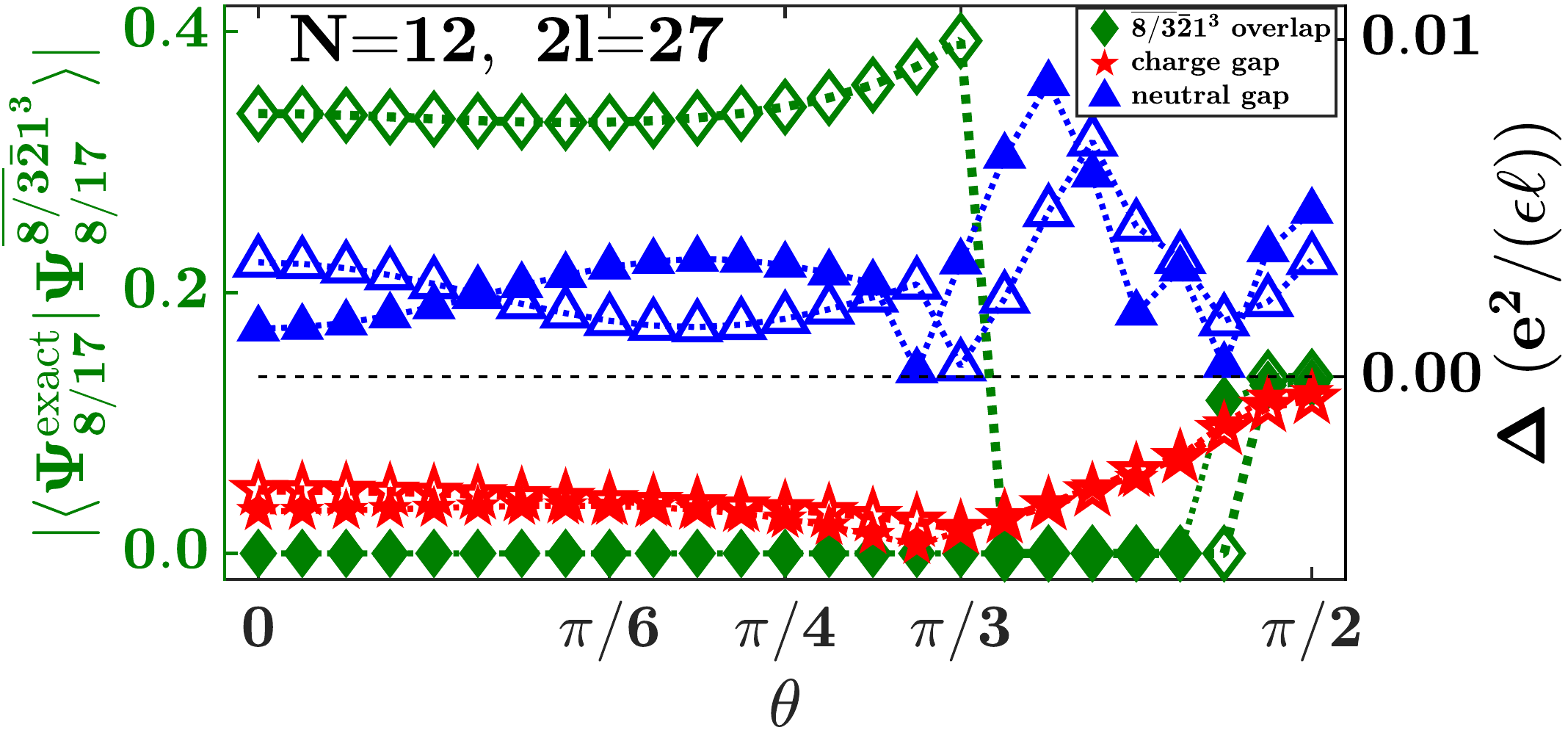} 
		\caption{(color online) Overlaps and gaps of the $\overline{(8/3)}\bar{2}1^{3}$ parton state with the exact Coulomb ground state in the zeroth Landau level of bilayer graphene at $\nu{=}8/17$ evaluated in the spherical geometry using the spherical (filled symbols) and disk (open symbols) pseudopotentials for $N{=}12$ electrons and $2l{=}27$ flux quanta as a function of the parameter $\theta$, which is related to the perpendicular magnetic field $B{=}93.06[\cot(\theta)]^{2}$ (see text).}
		\label{fig: gaps_overlaps_8_17_bar8_3bar2111_ZLL_BLG}
	\end{center}
\end{figure}

In Fig.~\ref{fig: gaps_overlaps_8_17_bar8_3bar2111_ZLL_BLG} we show overlaps of the $\overline{(8/3)}\bar{2}1^{3}$ with the exact Coulomb ground state in the ZLL of BLG for $N{=}12$. The $8/3$ state is constructed by brute-force projection to the LLL and we then use this $8/3$ state to construct the $8/17$ state given in Eq.~\eqref{eq: parton_8_17_bar_8_3_bar2111} following the method outlined in Refs.~\cite{Sreejith13, Balram20a} that involves expanding the state in the space of all $L{=}0$ states. The overlaps of the parton state with the exact state are small for all values of $\theta$ in the ZLL of BLG indicating that the parton state does not give a good microscopic description of the experimentally observed $8/17$ state~\cite{Zibrov17, Huang21, Assouline23, Hu23, Kumar24}. Nevertheless, the parton state may lie in the same topological phase as the experimentally realized $8/17$ state. For the parton state, the only system accessible to ED is $N{=}12$ since the next system size of $N{=}20$ is currently beyond our reach (Hilbert space dimension ${\simeq} 1.43 {\times} 10^{10}$). 

\begin{figure}[htpb]
	\begin{center}
		\includegraphics[width=0.49\textwidth,height=0.26\textwidth]{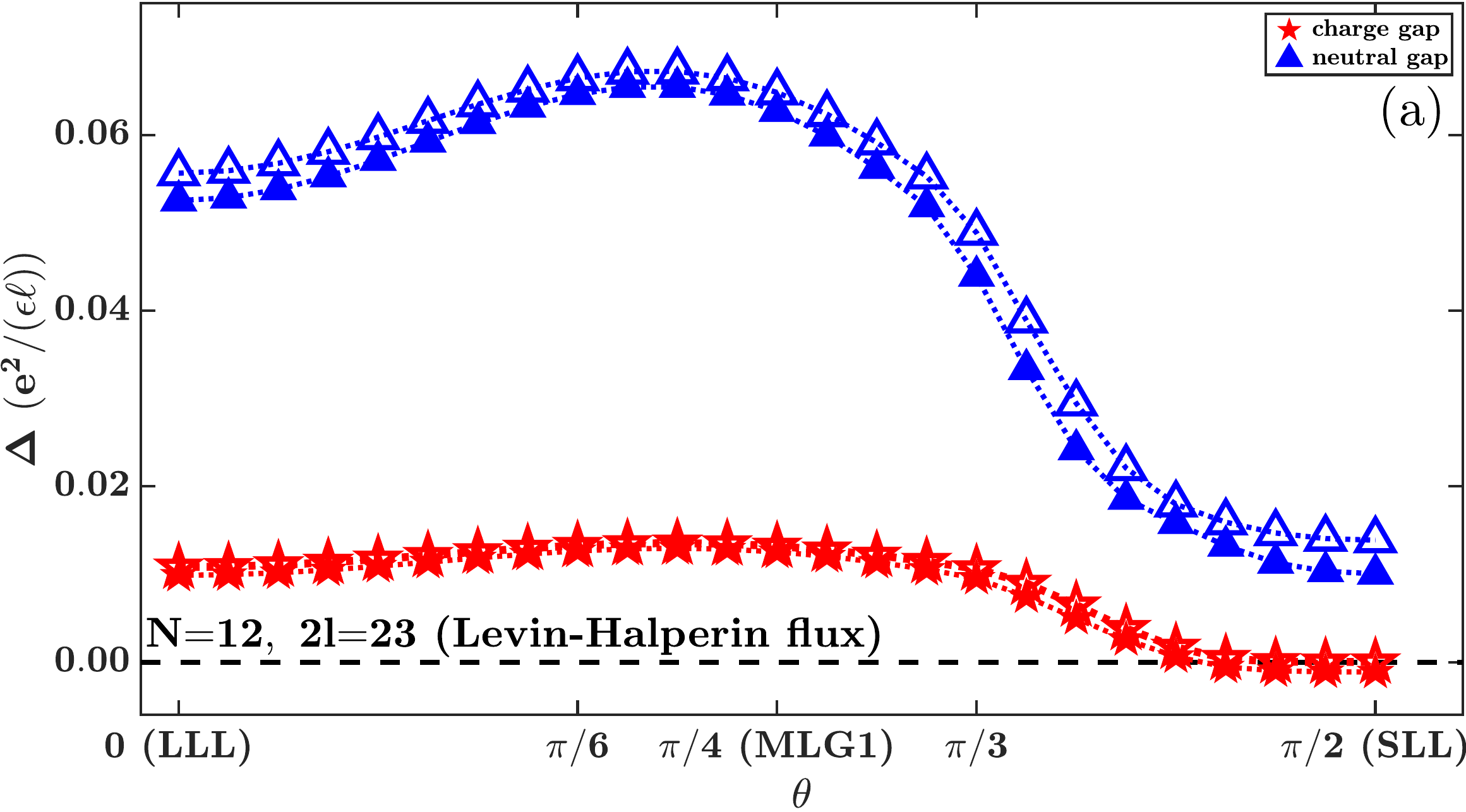}
            \includegraphics[width=0.49\textwidth,height=0.23\textwidth]{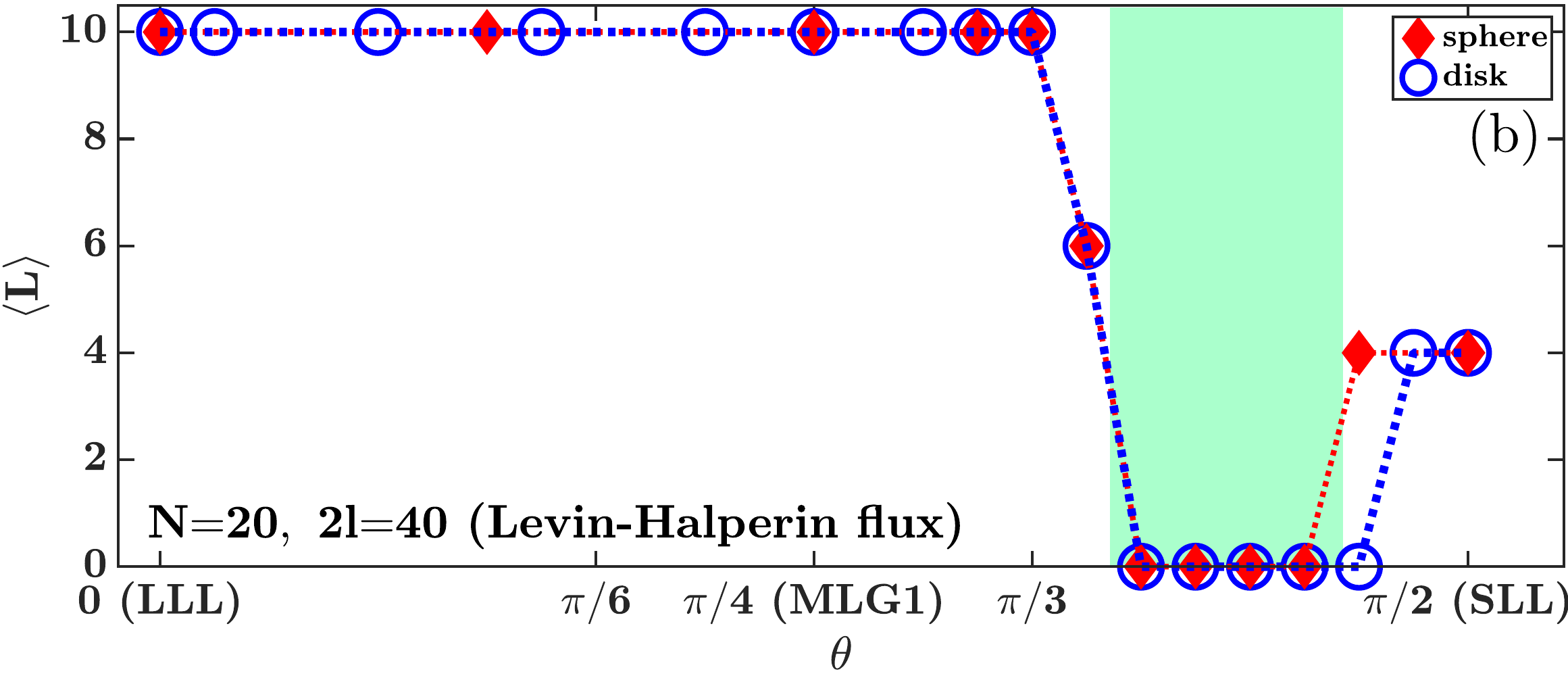}
		\caption{(color online) (a) Same as Fig.~\ref{fig: gaps_overlaps_8_17_bar8_3bar2111_ZLL_BLG} but at the LH flux $2l{=}23$ at $\nu{=}8/17$. (b) Total orbital angular momentum of the exact Coulomb ground state in the zeroth Landau level of bilayer graphene at $\nu{=}8/17$ at the LH flux for $N{=}20$ electrons.}
		\label{fig: gaps_8_17_LH_ZLL_BLG}
	\end{center}
\end{figure}

The LH state is not readily amenable to a numerical construction and we have not been able to obtain its wave function. However, we can carry out ED to obtain the ground state for $N{=}20$ (Hilbert space dimension ${\simeq} 1.38 {\times} 10^9$) at the shift corresponding to the LH state. In Fig.~\ref{fig: gaps_8_17_LH_ZLL_BLG}(b) we show the ground state $L$ as a function of $\theta$ for this system. There is a narrow region of parameter space of small to intermediate fields where the ground state at the LH shift in the ZLL of BLG for this system is uniform (Moore-Read quasiholes give a good description of the exact Coulomb ground state here~\cite{SM}.) while the exact SLL Coulomb ground state (that occurs at zero field) at this shift is not uniform. Note that for  $N{=}12$ at the LH shift, the ground state has $L{=}0$ for all $\theta$.

In Figs.~\ref{fig: gaps_overlaps_8_17_bar8_3bar2111_ZLL_BLG} and \ref{fig: gaps_8_17_LH_ZLL_BLG}(a), we show the charge and neutral gaps at the flux corresponding to the $\overline{(8/3)}\bar{2}1^{3}$ and the LH state in the ZLL of BLG for $N{=}12$ electrons. The charge gap is the energy to create a pair of fundamental qp-qh (the smallest charged qh in the $\overline{(8/3)}\bar{2}1^{3}$ and the LH state has charge $e/17$). The neutral gap is the energy difference between the ground state and the lowest-lying excitation. For both fluxes, the charge gap is not consistently positive and the neutral gap is low in the vicinity of the SLL point. Furthermore, the charge gap is much smaller than the neutral gap while in the thermodynamic limit, we expect the charge gap to be greater than the neutral gap. These results suggest the presence of strong finite-size effects (as has been routinely seen in numerics carried out in the SLL) and indicate that there could be aliasing effects in the spherical geometry~\cite{Ambrumenil89} where the same system can correspond to two different states. Thus, our numerical results are inconclusive in unambiguously identifying the topological order at 8/17. Next, we turn to effective field theory to make predictions that are experimentally measurable and can aid in identifying the underlying order at 8/17.

{\bf \em Effective field theory.---}The topological properties of the $\overline{(8/3)}\bar{2}1^{3}$ state can be read-off from its $K$ matrix~\cite{Wen91b,Wen92b,Wen95,Moore98}, charge vector $\vec{t}$ and the spin-vector $\vec{\mathfrak{s}}$~\cite{Wen92}. A straightforward extension of the derivation outlined in Ref.~\cite{Balram18a} for the $\bar{n}\bar{2}1^{3}$ states shows that the $K$, $\vec{t}$ and $\vec{\mathfrak{s}}$ for the $\overline{(8/3)}\bar{2}1^{3}$ are given by
\begin{equation}
K = 
\begin{pmatrix} 
     1 & 1 & 1 & 1  & 1 \\
     1 & -2 & -1 & -1  & 0 \\
     1 & -1 & -2 & -3  & 0  \\
     1 & -1 & -3 & -2  & 0  \\
     1 & 0 & 0 & 0  &  -2 \\
 \end{pmatrix},~
 \vec{t}= \begin{pmatrix} 
   1\\
   0\\
   0\\
   0\\
   0
 \end{pmatrix}~{\rm and}~
 \vec{\mathfrak{s}}= \begin{pmatrix} 
   -3/2 \\
   1 \\
   -1\\
   -1\\
   1
 \end{pmatrix}.
\label{eq: Kmatrix_parton_bar_8_3_bar2111}  
\end{equation}
The filling factor and shift are obtained from the $K$-matrix as~\cite{Wen95} $\nu {=} \vec{t}^{\rm T}{\cdot} K^{-1} {\cdot} \vec{t} {=} 8/17$ and $\mathcal{S}{=}(2/\nu) \left( \vec{t}^{\rm T}{\cdot} K^{-1} {\cdot} \vec{\mathfrak{s}} \right){=}{-}3/2$. These values are consistent with that determined from the microscopic wave function of the state given in Eq.~\eqref{eq: parton_8_17_bar_8_3_bar2111}. The ground-state degeneracy of the $\overline{(8/3)}\bar{2}1^{3}$ state on a manifold with genus $\mathfrak{g}$ is~\cite{Wen95} $\mathcal{D} {=} |{\rm Det}(K)|^{\mathfrak{g}} {=}34^{\mathfrak{g}}$. Therefore, this state presents an example of a single-component Abelian state at $\nu{=}p/q$ (with $p,q$ coprime) which has a ground-state degeneracy on the torus that is greater than $q$ (See Refs.~\cite{Balram20, Balram20b, Balram20a, Dora22} for other examples of such states.). The $K$ matrix of Eq.~\eqref{eq: Kmatrix_parton_bar_8_3_bar2111} has three negative and two positive eigenvalues resulting in a chiral central charge of ${-}1$. The presence of five edge states can be understood as follows: at the mean-field level, the parton theory results in a total of nine-edge modes four from the factor $\Phi_{2{+}2/3}$ [two each from $\nu{=}2$ and $\nu{=}2/3$], two from $\Phi_{2}$, and one each from each factor of $\Phi_{1}$. However, these edge modes are not all independent since the density fluctuations of the five partons have to be identified, which results in exactly four constraints and thereby leads to precisely five independent edge modes. For completeness, we point out that the edge structure of the LH states is worked out in Ref.~\cite{Levin09a}. 

{\bf \em Discussion.---}
In this section, we discuss various experimentally measurable properties of the $\overline{(8/3)}\bar{2}1^{3}$ ansatz that can reveal its underlying topological order. The smallest quasiparticle, generated by creating a hole in the third LL in the factor of $\Phi_{8/3}$, carries a charge of ${-}e/17$. A single quasiparticle of charge ${-}3e/17$, ${-}4e/17$ and ${-}8e/17$ can be produced by creating a hole in the LLL or the SLL in the factor of $\Phi_{8/3}$, a hole in the factor of $\Phi_{2}$ and a particle in the factor of $\Phi_{1}$ respectively. All the excitations of the $\overline{(8/3)}\bar{2}1^{3}$ state carry Abelian braiding statistics just like those of the LH state.

Due to the presence of the factors of $\overline{(8/3)}$ and $\bar{2}$, the $\overline{(8/3)}\bar{2}1^{3}$ state hosts upstream edge modes that can be detected experimentally~\cite{Bid10, Dolev11}. \emph{Assuming} a full equilibration of the edge modes, the thermal Hall conductance $\kappa_{xy}$ of the $\overline{(8/3)}\bar{2}1^{3}$ state is $\kappa_{xy}{=}({-}1)[\pi^2 k^{2}_{B}/(3h)T]$. In contrast, the $8/17$ LH state and the $8/17$ Jain states respectively have $\kappa^{8/17-{\rm LH}}_{xy}{=}0$~\cite{Levin09a} and $\kappa^{8/17-{\rm Jain}}_{xy}{=}8[\pi^2 k^{2}_{B}/(3h)T]$. Recently, the thermal Hall conductance of many quantum Hall states has been measured in GaAs~\cite{Banerjee17, Banerjee18} and monolayer graphene~\cite{Srivastav19}. The Hall viscosity $\eta_{H}$ of the $\overline{(8/3)}\bar{2}1^{3}$ state is also anticipated to be quantized~\cite{Read09}: $\eta_{H}{=}\hbar n_{{\rm 2D}} \mathcal{S}/4$, where $n_{{\rm 2D}}{=}(8/17)/(2\pi\ell^{2})$ is the density of the electrons and $\mathcal{S}{=}{-}3/2$ is the shift of the $\overline{(8/3)}\bar{2}1^{3}$ ansatz. For comparison, the $8/17$ LH state and the $8/17$ Jain states respectively have $\mathcal{S}^{8/17-{\rm LH}}{=}5/2$~\cite{Levin09a} and $\mathcal{S}^{8/17-{\rm Jain}}{=}10$. These results show that the $8/17$ LH, $8/17$ Jain and our proposed $\overline{(8/3)}\bar{2}1^{3}$ states are all topologically different from each other.

The $\overline{(8/3)}\bar{2}1^{3}$ state is the $m{=}1$ member of the $\overline{[2{+}2/(4m{-}1)]}\bar{2}1^{3}$ sequence which produces states at $\nu{=}8m/(16m{+}1)$ with shift $\mathcal{S}{=}{-}3/2$ and chiral central charge ${-}1$. A LH sequence~\cite{Levin09a} that is obtained by condensing qhs of the Pf also leads to states at the same set of fillings $\nu{=}8m/(16m{+}1)$ but with shift $\mathcal{S}{=}5/2$ and chiral central charge $0$. Thus, the parton and LH states are topologically distinct. The $\overline{(8/3)}\bar{2}1^{3}$ state is also the $m{=}1$ member of the $\overline{[2{+}(m{+}1)]/(2m{+}1)}\bar{2}1^{3}$ sequence which produces states at $26/55$ and $9/19$ for $m{=}2,3$. No signs of FQHE have been reported at these fractions.

Potentially, simpler-looking states such as $\overline{(4/3)}\bar{2}1^{3}$ at 4/7 or $\overline{(5/3)}\bar{2}1^{3}$ at 10/19, or $\overline{(7/3)}\bar{2}1^{3}$ at 14/29 could also be stabilized. Signatures of FQHE at $\nu{=}4/7$ in the ZLL of BLG have been seen in experiments~\cite{Huang21}. However, numerical calculations show that the ground state of the largest accessible system of $N{=}20$ electrons at the flux corresponding to $\overline{(4/3)}\bar{2}1^{3}$ is not uniform at the SLL Coulomb point. It turns out that the particle-hole conjugate of the $\bar{3}^{2}1^{3}$ state at 3/7~\cite{Faugno20a} is a better candidate in that the exact ground state is consistently incompressible at the corresponding shift for all accessible systems. Moreover, the overlap of the exact Coulomb ground state in the ZLL of BLG with the trial wave function is reasonably high~\cite{Balram20a, Balram21b}. No signs of FQHE have been seen in either the SLL of GaAs or in the ZLL of BLG at either 10/19 or 14/29. We mention here that the $\overline{(7/3)}\bar{2}1^{3}$ state at 14/29 is likely to be topologically identical to a LH state. It is the $m{=}2$ member of the $\overline{[2{+}1/(2m{-}1)]}\bar{2}1^{3}$ sequence that produces states at $\nu{=}(8m{-}2)/(16m{-}3)$ with shift $\mathcal{S}{=}{-}2$ and chiral central charge ${-}2$. A LH sequence~\cite{Levin09a} emanating from the aPf also leads to states at the same set of filling factors, that also carry the same chiral central charge and shift. Therefore, as a by-product of our construction, we found a parton sequence that produces states that likely lie in the same universality class as the LH states arising from the aPf. We have not been able to find parton states that lie in the same topological phase as the LH states built from the Pf. We leave a more detailed exploration of the relationship between parton and LH states to future work. 

If it turns out that the experimentally observed 8/17 state lies in the universality class of $\overline{(8/3)}\bar{2}1^{3}$ state, then it would call into question the idea that by looking at the nearby fractions one can tell if the even-denominator state seen at half-filling is in the Pf or the aPf universality class~\cite{Zibrov16, Huang21, Assouline23, Hu23, Singh23, Yutushui24, Zheltonozhskii24, Kumar24}. Finally, we mention the possibility of multi-component states at 8/17 where the different components could represent valley, spin, or orbital degree of freedom. The $\overline{(8/3)}\bar{2}1^{3}$ state readily admits the possibility of unpolarized states building on the partially-polarized and singlet-states at $\nu{=}8/3$ and the singlet state at $\nu{=}2$. It is possible that these states could potentially be relevant for certain interactions. Potentially all possible FQHE states lend themselves to a parton description where the partons themselves form simple IQHE or FQHE states.

{\bf \em Acknowledgments.---}
We acknowledge useful discussions with Maissam Barkeshli, William N. Faugno, Jainendra Jain, and Arkadius\'z W\'ojs. The work was made possible by financial support from the Science and Engineering Research Board (SERB) of the Department of Science and Technology (DST) via the Mathematical Research Impact Centric Support (MATRICS) Grant No. MTR/2023/000002. Computational portions of this research work were conducted using the Nandadevi supercomputer, which is maintained and supported by the Institute of Mathematical Science's High-Performance Computing Center. N.R. acknowledges support from the QuantERA II Programme which has received funding from the European Union’s Horizon 2020 research and innovation programme under Grant Agreement No 101017733. Numerical calculations were performed using the DiagHam package~\cite{diagham}.

\bibliography{biblio_FQHE}

\begin{thebibliography}{74}%
\makeatletter
\providecommand \@ifxundefined [1]{%
 \@ifx{#1\undefined}
}%
\providecommand \@ifnum [1]{%
 \ifnum #1\expandafter \@firstoftwo
 \else \expandafter \@secondoftwo
 \fi
}%
\providecommand \@ifx [1]{%
 \ifx #1\expandafter \@firstoftwo
 \else \expandafter \@secondoftwo
 \fi
}%
\providecommand \natexlab [1]{#1}%
\providecommand \enquote  [1]{``#1''}%
\providecommand \bibnamefont  [1]{#1}%
\providecommand \bibfnamefont [1]{#1}%
\providecommand \citenamefont [1]{#1}%
\providecommand \href@noop [0]{\@secondoftwo}%
\providecommand \href [0]{\begingroup \@sanitize@url \@href}%
\providecommand \@href[1]{\@@startlink{#1}\@@href}%
\providecommand \@@href[1]{\endgroup#1\@@endlink}%
\providecommand \@sanitize@url [0]{\catcode `\\12\catcode `\$12\catcode
  `\&12\catcode `\#12\catcode `\^12\catcode `\_12\catcode `\%12\relax}%
\providecommand \@@startlink[1]{}%
\providecommand \@@endlink[0]{}%
\providecommand \url  [0]{\begingroup\@sanitize@url \@url }%
\providecommand \@url [1]{\endgroup\@href {#1}{\urlprefix }}%
\providecommand \urlprefix  [0]{URL }%
\providecommand \Eprint [0]{\href }%
\providecommand \doibase [0]{https://doi.org/}%
\providecommand \selectlanguage [0]{\@gobble}%
\providecommand \bibinfo  [0]{\@secondoftwo}%
\providecommand \bibfield  [0]{\@secondoftwo}%
\providecommand \translation [1]{[#1]}%
\providecommand \BibitemOpen [0]{}%
\providecommand \bibitemStop [0]{}%
\providecommand \bibitemNoStop [0]{.\EOS\space}%
\providecommand \EOS [0]{\spacefactor3000\relax}%
\providecommand \BibitemShut  [1]{\csname bibitem#1\endcsname}%
\let\auto@bib@innerbib\@empty
\bibitem [{\citenamefont {Tsui}\ \emph {et~al.}(1982)\citenamefont {Tsui},
  \citenamefont {Stormer},\ and\ \citenamefont {Gossard}}]{Tsui82}%
  \BibitemOpen
  \bibfield  {author} {\bibinfo {author} {\bibfnamefont {D.~C.}\ \bibnamefont
  {Tsui}}, \bibinfo {author} {\bibfnamefont {H.~L.}\ \bibnamefont {Stormer}},\
  and\ \bibinfo {author} {\bibfnamefont {A.~C.}\ \bibnamefont {Gossard}},\
  }\bibfield  {title} {\bibinfo {title} {Two-dimensional magnetotransport in
  the extreme quantum limit},\ }\href
  {https://doi.org/10.1103/PhysRevLett.48.1559} {\bibfield  {journal} {\bibinfo
   {journal} {Phys. Rev. Lett.}\ }\textbf {\bibinfo {volume} {48}},\ \bibinfo
  {pages} {1559} (\bibinfo {year} {1982})}\BibitemShut {NoStop}%
\bibitem [{\citenamefont {Laughlin}(1983)}]{Laughlin83}%
  \BibitemOpen
  \bibfield  {author} {\bibinfo {author} {\bibfnamefont {R.~B.}\ \bibnamefont
  {Laughlin}},\ }\bibfield  {title} {\bibinfo {title} {Anomalous quantum {Hall}
  effect: An incompressible quantum fluid with fractionally charged
  excitations},\ }\href {https://doi.org/10.1103/PhysRevLett.50.1395}
  {\bibfield  {journal} {\bibinfo  {journal} {Phys. Rev. Lett.}\ }\textbf
  {\bibinfo {volume} {50}},\ \bibinfo {pages} {1395} (\bibinfo {year}
  {1983})}\BibitemShut {NoStop}%
\bibitem [{\citenamefont {Wen}(1995)}]{Wen95}%
  \BibitemOpen
  \bibfield  {author} {\bibinfo {author} {\bibfnamefont {X.-G.}\ \bibnamefont
  {Wen}},\ }\bibfield  {title} {\bibinfo {title} {Topological orders and edge
  excitations in fractional quantum {Hall} states},\ }\href
  {http://www.tandfonline.com/doi/abs/10.1080/00018739500101566} {\bibfield
  {journal} {\bibinfo  {journal} {Advances in Physics}\ }\textbf {\bibinfo
  {volume} {44}},\ \bibinfo {pages} {405} (\bibinfo {year} {1995})}\BibitemShut
  {NoStop}%
\bibitem [{\citenamefont {Haldane}(1983)}]{Haldane83}%
  \BibitemOpen
  \bibfield  {author} {\bibinfo {author} {\bibfnamefont {F.~D.~M.}\
  \bibnamefont {Haldane}},\ }\bibfield  {title} {\bibinfo {title} {Fractional
  quantization of the {Hall} effect: A hierarchy of incompressible quantum
  fluid states},\ }\href {https://doi.org/10.1103/PhysRevLett.51.605}
  {\bibfield  {journal} {\bibinfo  {journal} {Phys. Rev. Lett.}\ }\textbf
  {\bibinfo {volume} {51}},\ \bibinfo {pages} {605} (\bibinfo {year}
  {1983})}\BibitemShut {NoStop}%
\bibitem [{\citenamefont {Halperin}(1984)}]{Halperin84}%
  \BibitemOpen
  \bibfield  {author} {\bibinfo {author} {\bibfnamefont {B.~I.}\ \bibnamefont
  {Halperin}},\ }\bibfield  {title} {\bibinfo {title} {Statistics of
  quasiparticles and the hierarchy of fractional quantized {Hall} states},\
  }\href {https://doi.org/10.1103/PhysRevLett.52.1583} {\bibfield  {journal}
  {\bibinfo  {journal} {Phys. Rev. Lett.}\ }\textbf {\bibinfo {volume} {52}},\
  \bibinfo {pages} {1583} (\bibinfo {year} {1984})}\BibitemShut {NoStop}%
\bibitem [{\citenamefont {Jain}(1989{\natexlab{a}})}]{Jain89}%
  \BibitemOpen
  \bibfield  {author} {\bibinfo {author} {\bibfnamefont {J.~K.}\ \bibnamefont
  {Jain}},\ }\bibfield  {title} {\bibinfo {title} {Composite-fermion approach
  for the fractional quantum {Hall} effect},\ }\href
  {https://doi.org/10.1103/PhysRevLett.63.199} {\bibfield  {journal} {\bibinfo
  {journal} {Phys. Rev. Lett.}\ }\textbf {\bibinfo {volume} {63}},\ \bibinfo
  {pages} {199} (\bibinfo {year} {1989}{\natexlab{a}})}\BibitemShut {NoStop}%
\bibitem [{\citenamefont {Jain}(1989{\natexlab{b}})}]{Jain89b}%
  \BibitemOpen
  \bibfield  {author} {\bibinfo {author} {\bibfnamefont {J.~K.}\ \bibnamefont
  {Jain}},\ }\bibfield  {title} {\bibinfo {title} {Incompressible quantum
  {Hall} states},\ }\href {https://doi.org/10.1103/PhysRevB.40.8079} {\bibfield
   {journal} {\bibinfo  {journal} {Phys. Rev. B}\ }\textbf {\bibinfo {volume}
  {40}},\ \bibinfo {pages} {8079} (\bibinfo {year}
  {1989}{\natexlab{b}})}\BibitemShut {NoStop}%
\bibitem [{\citenamefont {Hansson}\ \emph {et~al.}(2017)\citenamefont
  {Hansson}, \citenamefont {Hermanns}, \citenamefont {Simon},\ and\
  \citenamefont {Viefers}}]{Hansson17}%
  \BibitemOpen
  \bibfield  {author} {\bibinfo {author} {\bibfnamefont {T.~H.}\ \bibnamefont
  {Hansson}}, \bibinfo {author} {\bibfnamefont {M.}~\bibnamefont {Hermanns}},
  \bibinfo {author} {\bibfnamefont {S.~H.}\ \bibnamefont {Simon}},\ and\
  \bibinfo {author} {\bibfnamefont {S.~F.}\ \bibnamefont {Viefers}},\
  }\bibfield  {title} {\bibinfo {title} {Quantum {Hall} physics: Hierarchies
  and conformal field theory techniques},\ }\href
  {https://doi.org/10.1103/RevModPhys.89.025005} {\bibfield  {journal}
  {\bibinfo  {journal} {Rev. Mod. Phys.}\ }\textbf {\bibinfo {volume} {89}},\
  \bibinfo {pages} {025005} (\bibinfo {year} {2017})}\BibitemShut {NoStop}%
\bibitem [{\citenamefont {Willett}\ \emph {et~al.}(1987)\citenamefont
  {Willett}, \citenamefont {Eisenstein}, \citenamefont {St\"ormer},
  \citenamefont {Tsui}, \citenamefont {Gossard},\ and\ \citenamefont
  {English}}]{Willett87}%
  \BibitemOpen
  \bibfield  {author} {\bibinfo {author} {\bibfnamefont {R.}~\bibnamefont
  {Willett}}, \bibinfo {author} {\bibfnamefont {J.~P.}\ \bibnamefont
  {Eisenstein}}, \bibinfo {author} {\bibfnamefont {H.~L.}\ \bibnamefont
  {St\"ormer}}, \bibinfo {author} {\bibfnamefont {D.~C.}\ \bibnamefont {Tsui}},
  \bibinfo {author} {\bibfnamefont {A.~C.}\ \bibnamefont {Gossard}},\ and\
  \bibinfo {author} {\bibfnamefont {J.~H.}\ \bibnamefont {English}},\
  }\bibfield  {title} {\bibinfo {title} {Observation of an even-denominator
  quantum number in the fractional quantum {Hall} effect},\ }\href
  {https://doi.org/10.1103/PhysRevLett.59.1776} {\bibfield  {journal} {\bibinfo
   {journal} {Phys. Rev. Lett.}\ }\textbf {\bibinfo {volume} {59}},\ \bibinfo
  {pages} {1776} (\bibinfo {year} {1987})}\BibitemShut {NoStop}%
\bibitem [{\citenamefont {Moore}\ and\ \citenamefont {Read}(1991)}]{Moore91}%
  \BibitemOpen
  \bibfield  {author} {\bibinfo {author} {\bibfnamefont {G.}~\bibnamefont
  {Moore}}\ and\ \bibinfo {author} {\bibfnamefont {N.}~\bibnamefont {Read}},\
  }\bibfield  {title} {\bibinfo {title} {Nonabelions in the fractional quantum
  {Hall} effect},\ }\href {https://doi.org/10.1016/0550-3213(91)90407-O}
  {\bibfield  {journal} {\bibinfo  {journal} {Nucl. Phys. B}\ }\textbf
  {\bibinfo {volume} {360}},\ \bibinfo {pages} {362 } (\bibinfo {year}
  {1991})}\BibitemShut {NoStop}%
\bibitem [{\citenamefont {Morf}\ \emph {et~al.}(2002)\citenamefont {Morf},
  \citenamefont {d'Ambrumenil},\ and\ \citenamefont {Das~Sarma}}]{Morf02}%
  \BibitemOpen
  \bibfield  {author} {\bibinfo {author} {\bibfnamefont {R.~H.}\ \bibnamefont
  {Morf}}, \bibinfo {author} {\bibfnamefont {N.}~\bibnamefont {d'Ambrumenil}},\
  and\ \bibinfo {author} {\bibfnamefont {S.}~\bibnamefont {Das~Sarma}},\
  }\bibfield  {title} {\bibinfo {title} {Excitation gaps in fractional quantum
  {Hall} states: An exact diagonalization study},\ }\href
  {https://doi.org/10.1103/PhysRevB.66.075408} {\bibfield  {journal} {\bibinfo
  {journal} {Phys. Rev. B}\ }\textbf {\bibinfo {volume} {66}},\ \bibinfo
  {pages} {075408} (\bibinfo {year} {2002})}\BibitemShut {NoStop}%
\bibitem [{\citenamefont {Balram}\ and\ \citenamefont
  {W\'ojs}(2020)}]{Balram20b}%
  \BibitemOpen
  \bibfield  {author} {\bibinfo {author} {\bibfnamefont {A.~C.}\ \bibnamefont
  {Balram}}\ and\ \bibinfo {author} {\bibfnamefont {A.}~\bibnamefont
  {W\'ojs}},\ }\bibfield  {title} {\bibinfo {title} {Fractional quantum {Hall}
  effect at $\ensuremath{\nu}=2+4/9$},\ }\href
  {https://doi.org/10.1103/PhysRevResearch.2.032035} {\bibfield  {journal}
  {\bibinfo  {journal} {Phys. Rev. Research}\ }\textbf {\bibinfo {volume}
  {2}},\ \bibinfo {pages} {032035} (\bibinfo {year} {2020})}\BibitemShut
  {NoStop}%
\bibitem [{\citenamefont {Read}\ and\ \citenamefont {Rezayi}(1999)}]{Read99}%
  \BibitemOpen
  \bibfield  {author} {\bibinfo {author} {\bibfnamefont {N.}~\bibnamefont
  {Read}}\ and\ \bibinfo {author} {\bibfnamefont {E.}~\bibnamefont {Rezayi}},\
  }\bibfield  {title} {\bibinfo {title} {Beyond paired quantum {Hall} states:
  Parafermions and incompressible states in the first excited {Landau} level},\
  }\href {https://doi.org/10.1103/PhysRevB.59.8084} {\bibfield  {journal}
  {\bibinfo  {journal} {Phys. Rev. B}\ }\textbf {\bibinfo {volume} {59}},\
  \bibinfo {pages} {8084} (\bibinfo {year} {1999})}\BibitemShut {NoStop}%
\bibitem [{\citenamefont {Rezayi}\ and\ \citenamefont {Read}(2009)}]{Rezayi09}%
  \BibitemOpen
  \bibfield  {author} {\bibinfo {author} {\bibfnamefont {E.~H.}\ \bibnamefont
  {Rezayi}}\ and\ \bibinfo {author} {\bibfnamefont {N.}~\bibnamefont {Read}},\
  }\bibfield  {title} {\bibinfo {title} {Non-abelian quantized {Hall} states of
  electrons at filling factors 12/5 and 13/5 in the first excited {Landau}
  level},\ }\href {https://doi.org/10.1103/PhysRevB.79.075306} {\bibfield
  {journal} {\bibinfo  {journal} {Phys. Rev. B}\ }\textbf {\bibinfo {volume}
  {79}},\ \bibinfo {pages} {075306} (\bibinfo {year} {2009})}\BibitemShut
  {NoStop}%
\bibitem [{\citenamefont {Peterson}\ \emph {et~al.}(2015)\citenamefont
  {Peterson}, \citenamefont {Wu}, \citenamefont {Cheng}, \citenamefont
  {Barkeshli}, \citenamefont {Wang},\ and\ \citenamefont
  {Das~Sarma}}]{Peterson15}%
  \BibitemOpen
  \bibfield  {author} {\bibinfo {author} {\bibfnamefont {M.~R.}\ \bibnamefont
  {Peterson}}, \bibinfo {author} {\bibfnamefont {Y.-L.}\ \bibnamefont {Wu}},
  \bibinfo {author} {\bibfnamefont {M.}~\bibnamefont {Cheng}}, \bibinfo
  {author} {\bibfnamefont {M.}~\bibnamefont {Barkeshli}}, \bibinfo {author}
  {\bibfnamefont {Z.}~\bibnamefont {Wang}},\ and\ \bibinfo {author}
  {\bibfnamefont {S.}~\bibnamefont {Das~Sarma}},\ }\bibfield  {title} {\bibinfo
  {title} {Abelian and non-abelian states in $\ensuremath{\nu}=2/3$ bilayer
  fractional quantum {Hall} systems},\ }\href
  {https://doi.org/10.1103/PhysRevB.92.035103} {\bibfield  {journal} {\bibinfo
  {journal} {Phys. Rev. B}\ }\textbf {\bibinfo {volume} {92}},\ \bibinfo
  {pages} {035103} (\bibinfo {year} {2015})}\BibitemShut {NoStop}%
\bibitem [{\citenamefont {Xia}\ \emph {et~al.}(2004)\citenamefont {Xia},
  \citenamefont {Pan}, \citenamefont {Vicente}, \citenamefont {Adams},
  \citenamefont {Sullivan}, \citenamefont {Stormer}, \citenamefont {Tsui},
  \citenamefont {Pfeiffer}, \citenamefont {Baldwin},\ and\ \citenamefont
  {West}}]{Xia04}%
  \BibitemOpen
  \bibfield  {author} {\bibinfo {author} {\bibfnamefont {J.~S.}\ \bibnamefont
  {Xia}}, \bibinfo {author} {\bibfnamefont {W.}~\bibnamefont {Pan}}, \bibinfo
  {author} {\bibfnamefont {C.~L.}\ \bibnamefont {Vicente}}, \bibinfo {author}
  {\bibfnamefont {E.~D.}\ \bibnamefont {Adams}}, \bibinfo {author}
  {\bibfnamefont {N.~S.}\ \bibnamefont {Sullivan}}, \bibinfo {author}
  {\bibfnamefont {H.~L.}\ \bibnamefont {Stormer}}, \bibinfo {author}
  {\bibfnamefont {D.~C.}\ \bibnamefont {Tsui}}, \bibinfo {author}
  {\bibfnamefont {L.~N.}\ \bibnamefont {Pfeiffer}}, \bibinfo {author}
  {\bibfnamefont {K.~W.}\ \bibnamefont {Baldwin}},\ and\ \bibinfo {author}
  {\bibfnamefont {K.~W.}\ \bibnamefont {West}},\ }\bibfield  {title} {\bibinfo
  {title} {Electron correlation in the second {Landau} level: A competition
  between many nearly degenerate quantum phases},\ }\href
  {https://doi.org/10.1103/PhysRevLett.93.176809} {\bibfield  {journal}
  {\bibinfo  {journal} {Phys. Rev. Lett.}\ }\textbf {\bibinfo {volume} {93}},\
  \bibinfo {pages} {176809} (\bibinfo {year} {2004})}\BibitemShut {NoStop}%
\bibitem [{\citenamefont {Kumar}\ \emph {et~al.}(2010)\citenamefont {Kumar},
  \citenamefont {Cs\'athy}, \citenamefont {Manfra}, \citenamefont {Pfeiffer},\
  and\ \citenamefont {West}}]{Kumar10}%
  \BibitemOpen
  \bibfield  {author} {\bibinfo {author} {\bibfnamefont {A.}~\bibnamefont
  {Kumar}}, \bibinfo {author} {\bibfnamefont {G.~A.}\ \bibnamefont {Cs\'athy}},
  \bibinfo {author} {\bibfnamefont {M.~J.}\ \bibnamefont {Manfra}}, \bibinfo
  {author} {\bibfnamefont {L.~N.}\ \bibnamefont {Pfeiffer}},\ and\ \bibinfo
  {author} {\bibfnamefont {K.~W.}\ \bibnamefont {West}},\ }\bibfield  {title}
  {\bibinfo {title} {Nonconventional odd-denominator fractional quantum {Hall}
  states in the second {Landau} level},\ }\href
  {https://doi.org/10.1103/PhysRevLett.105.246808} {\bibfield  {journal}
  {\bibinfo  {journal} {Phys. Rev. Lett.}\ }\textbf {\bibinfo {volume} {105}},\
  \bibinfo {pages} {246808} (\bibinfo {year} {2010})}\BibitemShut {NoStop}%
\bibitem [{\citenamefont {{Zibrov}}\ \emph {et~al.}(2017)\citenamefont
  {{Zibrov}}, \citenamefont {{Kometter}}, \citenamefont {{Zhou}}, \citenamefont
  {{Spanton}}, \citenamefont {{Taniguchi}}, \citenamefont {{Watanabe}},
  \citenamefont {{Zaletel}},\ and\ \citenamefont {{Young}}}]{Zibrov16}%
  \BibitemOpen
  \bibfield  {author} {\bibinfo {author} {\bibfnamefont {A.~A.}\ \bibnamefont
  {{Zibrov}}}, \bibinfo {author} {\bibfnamefont {C.~R.}\ \bibnamefont
  {{Kometter}}}, \bibinfo {author} {\bibfnamefont {H.}~\bibnamefont {{Zhou}}},
  \bibinfo {author} {\bibfnamefont {E.~M.}\ \bibnamefont {{Spanton}}}, \bibinfo
  {author} {\bibfnamefont {T.}~\bibnamefont {{Taniguchi}}}, \bibinfo {author}
  {\bibfnamefont {K.}~\bibnamefont {{Watanabe}}}, \bibinfo {author}
  {\bibfnamefont {M.~P.}\ \bibnamefont {{Zaletel}}},\ and\ \bibinfo {author}
  {\bibfnamefont {A.~F.}\ \bibnamefont {{Young}}},\ }\bibfield  {title}
  {\bibinfo {title} {Tunable interacting composite fermion phases in a
  half-filled bilayer-graphene {Landau} level},\ }\href
  {https://doi.org/10.1038/nature23893} {\bibfield  {journal} {\bibinfo
  {journal} {Nature}\ }\textbf {\bibinfo {volume} {549}},\ \bibinfo {pages}
  {360} (\bibinfo {year} {2017})}\BibitemShut {NoStop}%
\bibitem [{\citenamefont {Huang}\ \emph {et~al.}(2022)\citenamefont {Huang},
  \citenamefont {Fu}, \citenamefont {Hickey}, \citenamefont {Alem},
  \citenamefont {Lin}, \citenamefont {Watanabe}, \citenamefont {Taniguchi},\
  and\ \citenamefont {Zhu}}]{Huang21}%
  \BibitemOpen
  \bibfield  {author} {\bibinfo {author} {\bibfnamefont {K.}~\bibnamefont
  {Huang}}, \bibinfo {author} {\bibfnamefont {H.}~\bibnamefont {Fu}}, \bibinfo
  {author} {\bibfnamefont {D.~R.}\ \bibnamefont {Hickey}}, \bibinfo {author}
  {\bibfnamefont {N.}~\bibnamefont {Alem}}, \bibinfo {author} {\bibfnamefont
  {X.}~\bibnamefont {Lin}}, \bibinfo {author} {\bibfnamefont {K.}~\bibnamefont
  {Watanabe}}, \bibinfo {author} {\bibfnamefont {T.}~\bibnamefont
  {Taniguchi}},\ and\ \bibinfo {author} {\bibfnamefont {J.}~\bibnamefont
  {Zhu}},\ }\bibfield  {title} {\bibinfo {title} {Valley isospin controlled
  fractional quantum {Hall} states in bilayer graphene},\ }\href
  {https://doi.org/10.1103/PhysRevX.12.031019} {\bibfield  {journal} {\bibinfo
  {journal} {Phys. Rev. X}\ }\textbf {\bibinfo {volume} {12}},\ \bibinfo
  {pages} {031019} (\bibinfo {year} {2022})}\BibitemShut {NoStop}%
\bibitem [{\citenamefont {Assouline}\ \emph {et~al.}(2024)\citenamefont
  {Assouline}, \citenamefont {Wang}, \citenamefont {Zhou}, \citenamefont
  {Cohen}, \citenamefont {Yang}, \citenamefont {Zhang}, \citenamefont
  {Taniguchi}, \citenamefont {Watanabe}, \citenamefont {Mong}, \citenamefont
  {Zaletel},\ and\ \citenamefont {Young}}]{Assouline23}%
  \BibitemOpen
  \bibfield  {author} {\bibinfo {author} {\bibfnamefont {A.}~\bibnamefont
  {Assouline}}, \bibinfo {author} {\bibfnamefont {T.}~\bibnamefont {Wang}},
  \bibinfo {author} {\bibfnamefont {H.}~\bibnamefont {Zhou}}, \bibinfo {author}
  {\bibfnamefont {L.~A.}\ \bibnamefont {Cohen}}, \bibinfo {author}
  {\bibfnamefont {F.}~\bibnamefont {Yang}}, \bibinfo {author} {\bibfnamefont
  {R.}~\bibnamefont {Zhang}}, \bibinfo {author} {\bibfnamefont
  {T.}~\bibnamefont {Taniguchi}}, \bibinfo {author} {\bibfnamefont
  {K.}~\bibnamefont {Watanabe}}, \bibinfo {author} {\bibfnamefont {R.~S.~K.}\
  \bibnamefont {Mong}}, \bibinfo {author} {\bibfnamefont {M.~P.}\ \bibnamefont
  {Zaletel}},\ and\ \bibinfo {author} {\bibfnamefont {A.~F.}\ \bibnamefont
  {Young}},\ }\bibfield  {title} {\bibinfo {title} {Energy gap of the
  even-denominator fractional quantum {Hall} state in bilayer graphene},\
  }\href {https://doi.org/10.1103/PhysRevLett.132.046603} {\bibfield  {journal}
  {\bibinfo  {journal} {Phys. Rev. Lett.}\ }\textbf {\bibinfo {volume} {132}},\
  \bibinfo {pages} {046603} (\bibinfo {year} {2024})}\BibitemShut {NoStop}%
\bibitem [{\citenamefont {Hu}\ \emph {et~al.}(2023)\citenamefont {Hu},
  \citenamefont {Tsui}, \citenamefont {He}, \citenamefont {Kamber},
  \citenamefont {Wang}, \citenamefont {Mohammadi}, \citenamefont {Watanabe},
  \citenamefont {Taniguchi}, \citenamefont {Papic}, \citenamefont {Zaletel},\
  and\ \citenamefont {Yazdani}}]{Hu23}%
  \BibitemOpen
  \bibfield  {author} {\bibinfo {author} {\bibfnamefont {Y.}~\bibnamefont
  {Hu}}, \bibinfo {author} {\bibfnamefont {Y.-C.}\ \bibnamefont {Tsui}},
  \bibinfo {author} {\bibfnamefont {M.}~\bibnamefont {He}}, \bibinfo {author}
  {\bibfnamefont {U.}~\bibnamefont {Kamber}}, \bibinfo {author} {\bibfnamefont
  {T.}~\bibnamefont {Wang}}, \bibinfo {author} {\bibfnamefont {A.~S.}\
  \bibnamefont {Mohammadi}}, \bibinfo {author} {\bibfnamefont {K.}~\bibnamefont
  {Watanabe}}, \bibinfo {author} {\bibfnamefont {T.}~\bibnamefont {Taniguchi}},
  \bibinfo {author} {\bibfnamefont {Z.}~\bibnamefont {Papic}}, \bibinfo
  {author} {\bibfnamefont {M.~P.}\ \bibnamefont {Zaletel}},\ and\ \bibinfo
  {author} {\bibfnamefont {A.}~\bibnamefont {Yazdani}},\ }\href@noop {}
  {\bibinfo {title} {High-resolution tunneling spectroscopy of fractional
  quantum {Hall} states}} (\bibinfo {year} {2023}),\ \Eprint
  {https://arxiv.org/abs/2308.05789} {arXiv:2308.05789 [cond-mat.mes-hall]}
  \BibitemShut {NoStop}%
\bibitem [{\citenamefont {Kumar}\ \emph {et~al.}(2024)\citenamefont {Kumar},
  \citenamefont {Haug}, \citenamefont {Kim}, \citenamefont {Yutushui},
  \citenamefont {Khudiakov}, \citenamefont {Bhardwaj}, \citenamefont {Ilin},
  \citenamefont {Watanabe}, \citenamefont {Taniguchi}, \citenamefont {Mross},\
  and\ \citenamefont {Ronen}}]{Kumar24}%
  \BibitemOpen
  \bibfield  {author} {\bibinfo {author} {\bibfnamefont {R.}~\bibnamefont
  {Kumar}}, \bibinfo {author} {\bibfnamefont {A.}~\bibnamefont {Haug}},
  \bibinfo {author} {\bibfnamefont {J.}~\bibnamefont {Kim}}, \bibinfo {author}
  {\bibfnamefont {M.}~\bibnamefont {Yutushui}}, \bibinfo {author}
  {\bibfnamefont {K.}~\bibnamefont {Khudiakov}}, \bibinfo {author}
  {\bibfnamefont {V.}~\bibnamefont {Bhardwaj}}, \bibinfo {author}
  {\bibfnamefont {A.}~\bibnamefont {Ilin}}, \bibinfo {author} {\bibfnamefont
  {K.}~\bibnamefont {Watanabe}}, \bibinfo {author} {\bibfnamefont
  {T.}~\bibnamefont {Taniguchi}}, \bibinfo {author} {\bibfnamefont {D.~F.}\
  \bibnamefont {Mross}},\ and\ \bibinfo {author} {\bibfnamefont
  {Y.}~\bibnamefont {Ronen}},\ }\href@noop {} {\bibinfo {title} {Quarter- and
  half-filled quantum {Hall} states and their competing interactions in bilayer
  graphene}} (\bibinfo {year} {2024}),\ \Eprint
  {https://arxiv.org/abs/2405.19405} {arXiv:2405.19405 [cond-mat.mes-hall]}
  \BibitemShut {NoStop}%
\bibitem [{\citenamefont {Levin}\ and\ \citenamefont
  {Halperin}(2009)}]{Levin09a}%
  \BibitemOpen
  \bibfield  {author} {\bibinfo {author} {\bibfnamefont {M.}~\bibnamefont
  {Levin}}\ and\ \bibinfo {author} {\bibfnamefont {B.~I.}\ \bibnamefont
  {Halperin}},\ }\bibfield  {title} {\bibinfo {title} {Collective states of
  non-abelian quasiparticles in a magnetic field},\ }\href
  {https://doi.org/10.1103/PhysRevB.79.205301} {\bibfield  {journal} {\bibinfo
  {journal} {Phys. Rev. B}\ }\textbf {\bibinfo {volume} {79}},\ \bibinfo
  {pages} {205301} (\bibinfo {year} {2009})}\BibitemShut {NoStop}%
\bibitem [{\citenamefont {Levin}\ \emph {et~al.}(2007)\citenamefont {Levin},
  \citenamefont {Halperin},\ and\ \citenamefont {Rosenow}}]{Levin07}%
  \BibitemOpen
  \bibfield  {author} {\bibinfo {author} {\bibfnamefont {M.}~\bibnamefont
  {Levin}}, \bibinfo {author} {\bibfnamefont {B.~I.}\ \bibnamefont
  {Halperin}},\ and\ \bibinfo {author} {\bibfnamefont {B.}~\bibnamefont
  {Rosenow}},\ }\bibfield  {title} {\bibinfo {title} {Particle-hole symmetry
  and the {Pfaffian} state},\ }\href
  {https://doi.org/10.1103/PhysRevLett.99.236806} {\bibfield  {journal}
  {\bibinfo  {journal} {Phys. Rev. Lett.}\ }\textbf {\bibinfo {volume} {99}},\
  \bibinfo {pages} {236806} (\bibinfo {year} {2007})}\BibitemShut {NoStop}%
\bibitem [{\citenamefont {Lee}\ \emph {et~al.}(2007)\citenamefont {Lee},
  \citenamefont {Ryu}, \citenamefont {Nayak},\ and\ \citenamefont
  {Fisher}}]{Lee07}%
  \BibitemOpen
  \bibfield  {author} {\bibinfo {author} {\bibfnamefont {S.-S.}\ \bibnamefont
  {Lee}}, \bibinfo {author} {\bibfnamefont {S.}~\bibnamefont {Ryu}}, \bibinfo
  {author} {\bibfnamefont {C.}~\bibnamefont {Nayak}},\ and\ \bibinfo {author}
  {\bibfnamefont {M.~P.~A.}\ \bibnamefont {Fisher}},\ }\bibfield  {title}
  {\bibinfo {title} {Particle-hole symmetry and the $\nu=5/2$ quantum {Hall}
  state},\ }\href {https://doi.org/10.1103/PhysRevLett.99.236807} {\bibfield
  {journal} {\bibinfo  {journal} {Phys. Rev. Lett.}\ }\textbf {\bibinfo
  {volume} {99}},\ \bibinfo {pages} {236807} (\bibinfo {year}
  {2007})}\BibitemShut {NoStop}%
\bibitem [{\citenamefont {Singh}\ \emph {et~al.}(2024)\citenamefont {Singh},
  \citenamefont {Wang}, \citenamefont {Tai}, \citenamefont {Calhoun},
  \citenamefont {Villegas~Rosales}, \citenamefont {Madathil}, \citenamefont
  {Gupta}, \citenamefont {Baldwin}, \citenamefont {Pfeiffer},\ and\
  \citenamefont {Shayegan}}]{Singh23}%
  \BibitemOpen
  \bibfield  {author} {\bibinfo {author} {\bibfnamefont {S.~K.}\ \bibnamefont
  {Singh}}, \bibinfo {author} {\bibfnamefont {C.}~\bibnamefont {Wang}},
  \bibinfo {author} {\bibfnamefont {C.~T.}\ \bibnamefont {Tai}}, \bibinfo
  {author} {\bibfnamefont {C.~S.}\ \bibnamefont {Calhoun}}, \bibinfo {author}
  {\bibfnamefont {K.~A.}\ \bibnamefont {Villegas~Rosales}}, \bibinfo {author}
  {\bibfnamefont {P.~T.}\ \bibnamefont {Madathil}}, \bibinfo {author}
  {\bibfnamefont {A.}~\bibnamefont {Gupta}}, \bibinfo {author} {\bibfnamefont
  {K.~W.}\ \bibnamefont {Baldwin}}, \bibinfo {author} {\bibfnamefont {L.~N.}\
  \bibnamefont {Pfeiffer}},\ and\ \bibinfo {author} {\bibfnamefont
  {M.}~\bibnamefont {Shayegan}},\ }\bibfield  {title} {\bibinfo {title}
  {Topological phase transition between {Jain} states and daughter states of
  the $\nu${\thinspace}={\thinspace}1/2 fractional quantum {Hall} state},\
  }\bibfield  {journal} {\bibinfo  {journal} {Nature Physics}\ }\href
  {https://doi.org/10.1038/s41567-024-02517-w} {10.1038/s41567-024-02517-w}
  (\bibinfo {year} {2024})\BibitemShut {NoStop}%
\bibitem [{\citenamefont {Zhao}\ \emph {et~al.}(2021)\citenamefont {Zhao},
  \citenamefont {Faugno}, \citenamefont {Pu}, \citenamefont {Balram},\ and\
  \citenamefont {Jain}}]{Zhao21}%
  \BibitemOpen
  \bibfield  {author} {\bibinfo {author} {\bibfnamefont {T.}~\bibnamefont
  {Zhao}}, \bibinfo {author} {\bibfnamefont {W.~N.}\ \bibnamefont {Faugno}},
  \bibinfo {author} {\bibfnamefont {S.}~\bibnamefont {Pu}}, \bibinfo {author}
  {\bibfnamefont {A.~C.}\ \bibnamefont {Balram}},\ and\ \bibinfo {author}
  {\bibfnamefont {J.~K.}\ \bibnamefont {Jain}},\ }\bibfield  {title} {\bibinfo
  {title} {Origin of the $\ensuremath{\nu}=1/2$ fractional quantum {Hall}
  effect in wide quantum wells},\ }\href
  {https://doi.org/10.1103/PhysRevB.103.155306} {\bibfield  {journal} {\bibinfo
   {journal} {Phys. Rev. B}\ }\textbf {\bibinfo {volume} {103}},\ \bibinfo
  {pages} {155306} (\bibinfo {year} {2021})}\BibitemShut {NoStop}%
\bibitem [{\citenamefont {Sharma}\ \emph {et~al.}(2024)\citenamefont {Sharma},
  \citenamefont {Balram},\ and\ \citenamefont {Jain}}]{Sharma23}%
  \BibitemOpen
  \bibfield  {author} {\bibinfo {author} {\bibfnamefont {A.}~\bibnamefont
  {Sharma}}, \bibinfo {author} {\bibfnamefont {A.~C.}\ \bibnamefont {Balram}},\
  and\ \bibinfo {author} {\bibfnamefont {J.~K.}\ \bibnamefont {Jain}},\
  }\bibfield  {title} {\bibinfo {title} {Composite-fermion pairing at
  half-filled and quarter-filled lowest {Landau} level},\ }\href
  {https://doi.org/10.1103/PhysRevB.109.035306} {\bibfield  {journal} {\bibinfo
   {journal} {Phys. Rev. B}\ }\textbf {\bibinfo {volume} {109}},\ \bibinfo
  {pages} {035306} (\bibinfo {year} {2024})}\BibitemShut {NoStop}%
\bibitem [{\citenamefont {Balram}\ \emph
  {et~al.}(2018{\natexlab{a}})\citenamefont {Balram}, \citenamefont
  {Barkeshli},\ and\ \citenamefont {Rudner}}]{Balram18}%
  \BibitemOpen
  \bibfield  {author} {\bibinfo {author} {\bibfnamefont {A.~C.}\ \bibnamefont
  {Balram}}, \bibinfo {author} {\bibfnamefont {M.}~\bibnamefont {Barkeshli}},\
  and\ \bibinfo {author} {\bibfnamefont {M.~S.}\ \bibnamefont {Rudner}},\
  }\bibfield  {title} {\bibinfo {title} {Parton construction of a wave function
  in the anti-{Pfaffian} phase},\ }\href
  {https://doi.org/10.1103/PhysRevB.98.035127} {\bibfield  {journal} {\bibinfo
  {journal} {Phys. Rev. B}\ }\textbf {\bibinfo {volume} {98}},\ \bibinfo
  {pages} {035127} (\bibinfo {year} {2018}{\natexlab{a}})}\BibitemShut
  {NoStop}%
\bibitem [{\citenamefont {Balram}\ \emph {et~al.}(2019)\citenamefont {Balram},
  \citenamefont {Barkeshli},\ and\ \citenamefont {Rudner}}]{Balram19}%
  \BibitemOpen
  \bibfield  {author} {\bibinfo {author} {\bibfnamefont {A.~C.}\ \bibnamefont
  {Balram}}, \bibinfo {author} {\bibfnamefont {M.}~\bibnamefont {Barkeshli}},\
  and\ \bibinfo {author} {\bibfnamefont {M.~S.}\ \bibnamefont {Rudner}},\
  }\bibfield  {title} {\bibinfo {title} {Parton construction of
  particle-hole-conjugate {Read}-{Rezayi} parafermion fractional quantum {Hall}
  states and beyond},\ }\href {https://doi.org/10.1103/PhysRevB.99.241108}
  {\bibfield  {journal} {\bibinfo  {journal} {Phys. Rev. B}\ }\textbf {\bibinfo
  {volume} {99}},\ \bibinfo {pages} {241108} (\bibinfo {year}
  {2019})}\BibitemShut {NoStop}%
\bibitem [{\citenamefont {Balram}(2021{\natexlab{a}})}]{Balram20a}%
  \BibitemOpen
  \bibfield  {author} {\bibinfo {author} {\bibfnamefont {A.~C.}\ \bibnamefont
  {Balram}},\ }\bibfield  {title} {\bibinfo {title} {A non-abelian parton state
  for the $\ensuremath{\nu}=2+3/8$ fractional quantum {Hall} effect},\ }\href
  {https://doi.org/10.21468/SciPostPhys.10.4.083} {\bibfield  {journal}
  {\bibinfo  {journal} {SciPost Phys.}\ }\textbf {\bibinfo {volume} {10}},\
  \bibinfo {pages} {83} (\bibinfo {year} {2021}{\natexlab{a}})}\BibitemShut
  {NoStop}%
\bibitem [{\citenamefont {Bose}\ and\ \citenamefont {Balram}(2023)}]{Bose23}%
  \BibitemOpen
  \bibfield  {author} {\bibinfo {author} {\bibfnamefont {K.}~\bibnamefont
  {Bose}}\ and\ \bibinfo {author} {\bibfnamefont {A.~C.}\ \bibnamefont
  {Balram}},\ }\bibfield  {title} {\bibinfo {title} {Prediction of non-abelian
  fractional quantum {Hall} effect at $\ensuremath{\nu}=2+\frac{4}{11}$},\
  }\href {https://doi.org/10.1103/PhysRevB.107.235111} {\bibfield  {journal}
  {\bibinfo  {journal} {Phys. Rev. B}\ }\textbf {\bibinfo {volume} {107}},\
  \bibinfo {pages} {235111} (\bibinfo {year} {2023})}\BibitemShut {NoStop}%
\bibitem [{\citenamefont {Wen}\ and\ \citenamefont {Zee}(1992)}]{Wen92}%
  \BibitemOpen
  \bibfield  {author} {\bibinfo {author} {\bibfnamefont {X.~G.}\ \bibnamefont
  {Wen}}\ and\ \bibinfo {author} {\bibfnamefont {A.}~\bibnamefont {Zee}},\
  }\bibfield  {title} {\bibinfo {title} {Shift and spin vector: New topological
  quantum numbers for the {Hall} fluids},\ }\href
  {https://doi.org/10.1103/PhysRevLett.69.953} {\bibfield  {journal} {\bibinfo
  {journal} {Phys. Rev. Lett.}\ }\textbf {\bibinfo {volume} {69}},\ \bibinfo
  {pages} {953} (\bibinfo {year} {1992})}\BibitemShut {NoStop}%
\bibitem [{\citenamefont {Mukherjee}\ \emph {et~al.}(2014)\citenamefont
  {Mukherjee}, \citenamefont {Mandal}, \citenamefont {Wu}, \citenamefont
  {W\'ojs},\ and\ \citenamefont {Jain}}]{Mukherjee14}%
  \BibitemOpen
  \bibfield  {author} {\bibinfo {author} {\bibfnamefont {S.}~\bibnamefont
  {Mukherjee}}, \bibinfo {author} {\bibfnamefont {S.~S.}\ \bibnamefont
  {Mandal}}, \bibinfo {author} {\bibfnamefont {Y.-H.}\ \bibnamefont {Wu}},
  \bibinfo {author} {\bibfnamefont {A.}~\bibnamefont {W\'ojs}},\ and\ \bibinfo
  {author} {\bibfnamefont {J.~K.}\ \bibnamefont {Jain}},\ }\bibfield  {title}
  {\bibinfo {title} {Enigmatic $4/11$ state: A prototype for unconventional
  fractional quantum {Hall} effect},\ }\href
  {https://doi.org/10.1103/PhysRevLett.112.016801} {\bibfield  {journal}
  {\bibinfo  {journal} {Phys. Rev. Lett.}\ }\textbf {\bibinfo {volume} {112}},\
  \bibinfo {pages} {016801} (\bibinfo {year} {2014})}\BibitemShut {NoStop}%
\bibitem [{\citenamefont {Balram}\ \emph
  {et~al.}(2015{\natexlab{a}})\citenamefont {Balram}, \citenamefont {T\"oke},
  \citenamefont {W\'ojs},\ and\ \citenamefont {Jain}}]{Balram15a}%
  \BibitemOpen
  \bibfield  {author} {\bibinfo {author} {\bibfnamefont {A.~C.}\ \bibnamefont
  {Balram}}, \bibinfo {author} {\bibfnamefont {C.}~\bibnamefont {T\"oke}},
  \bibinfo {author} {\bibfnamefont {A.}~\bibnamefont {W\'ojs}},\ and\ \bibinfo
  {author} {\bibfnamefont {J.~K.}\ \bibnamefont {Jain}},\ }\bibfield  {title}
  {\bibinfo {title} {Fractional quantum {Hall} effect in graphene: Quantitative
  comparison between theory and experiment},\ }\href
  {https://doi.org/10.1103/PhysRevB.92.075410} {\bibfield  {journal} {\bibinfo
  {journal} {Phys. Rev. B}\ }\textbf {\bibinfo {volume} {92}},\ \bibinfo
  {pages} {075410} (\bibinfo {year} {2015}{\natexlab{a}})}\BibitemShut
  {NoStop}%
\bibitem [{\citenamefont {Balram}(2016)}]{Balram16c}%
  \BibitemOpen
  \bibfield  {author} {\bibinfo {author} {\bibfnamefont {A.~C.}\ \bibnamefont
  {Balram}},\ }\bibfield  {title} {\bibinfo {title} {Interacting composite
  fermions: Nature of the 4/5, 5/7, 6/7, and 6/17 fractional quantum {Hall}
  states},\ }\href {https://doi.org/10.1103/PhysRevB.94.165303} {\bibfield
  {journal} {\bibinfo  {journal} {Phys. Rev. B}\ }\textbf {\bibinfo {volume}
  {94}},\ \bibinfo {pages} {165303} (\bibinfo {year} {2016})}\BibitemShut
  {NoStop}%
\bibitem [{\citenamefont {Wu}\ \emph {et~al.}(2017)\citenamefont {Wu},
  \citenamefont {Shi},\ and\ \citenamefont {Jain}}]{Wu17}%
  \BibitemOpen
  \bibfield  {author} {\bibinfo {author} {\bibfnamefont {Y.}~\bibnamefont
  {Wu}}, \bibinfo {author} {\bibfnamefont {T.}~\bibnamefont {Shi}},\ and\
  \bibinfo {author} {\bibfnamefont {J.~K.}\ \bibnamefont {Jain}},\ }\bibfield
  {title} {\bibinfo {title} {Non-abelian parton fractional quantum {Hall}
  effect in multilayer graphene},\ }\href
  {https://doi.org/10.1021/acs.nanolett.7b01080} {\bibfield  {journal}
  {\bibinfo  {journal} {Nano Letters}\ }\textbf {\bibinfo {volume} {17}},\
  \bibinfo {pages} {4643} (\bibinfo {year} {2017})}\BibitemShut {NoStop}%
\bibitem [{\citenamefont {Kim}\ \emph {et~al.}(2019)\citenamefont {Kim},
  \citenamefont {Balram}, \citenamefont {Taniguchi}, \citenamefont {Watanabe},
  \citenamefont {Jain},\ and\ \citenamefont {Smet}}]{Kim19}%
  \BibitemOpen
  \bibfield  {author} {\bibinfo {author} {\bibfnamefont {Y.}~\bibnamefont
  {Kim}}, \bibinfo {author} {\bibfnamefont {A.~C.}\ \bibnamefont {Balram}},
  \bibinfo {author} {\bibfnamefont {T.}~\bibnamefont {Taniguchi}}, \bibinfo
  {author} {\bibfnamefont {K.}~\bibnamefont {Watanabe}}, \bibinfo {author}
  {\bibfnamefont {J.~K.}\ \bibnamefont {Jain}},\ and\ \bibinfo {author}
  {\bibfnamefont {J.~H.}\ \bibnamefont {Smet}},\ }\bibfield  {title} {\bibinfo
  {title} {Even denominator fractional quantum {Hall} states in higher {Landau}
  levels of graphene},\ }\href {https://doi.org/10.1038/s41567-018-0355-x}
  {\bibfield  {journal} {\bibinfo  {journal} {Nature Physics}\ }\textbf
  {\bibinfo {volume} {15}},\ \bibinfo {pages} {154} (\bibinfo {year}
  {2019})}\BibitemShut {NoStop}%
\bibitem [{\citenamefont {Faugno}\ \emph {et~al.}(2019)\citenamefont {Faugno},
  \citenamefont {Balram}, \citenamefont {Barkeshli},\ and\ \citenamefont
  {Jain}}]{Faugno19}%
  \BibitemOpen
  \bibfield  {author} {\bibinfo {author} {\bibfnamefont {W.~N.}\ \bibnamefont
  {Faugno}}, \bibinfo {author} {\bibfnamefont {A.~C.}\ \bibnamefont {Balram}},
  \bibinfo {author} {\bibfnamefont {M.}~\bibnamefont {Barkeshli}},\ and\
  \bibinfo {author} {\bibfnamefont {J.~K.}\ \bibnamefont {Jain}},\ }\bibfield
  {title} {\bibinfo {title} {Prediction of a non-{Abelian} fractional quantum
  {Hall} state with $f$-wave pairing of composite fermions in wide quantum
  wells},\ }\href {https://doi.org/10.1103/PhysRevLett.123.016802} {\bibfield
  {journal} {\bibinfo  {journal} {Phys. Rev. Lett.}\ }\textbf {\bibinfo
  {volume} {123}},\ \bibinfo {pages} {016802} (\bibinfo {year}
  {2019})}\BibitemShut {NoStop}%
\bibitem [{\citenamefont {Faugno}\ \emph {et~al.}(2020)\citenamefont {Faugno},
  \citenamefont {Jain},\ and\ \citenamefont {Balram}}]{Faugno20a}%
  \BibitemOpen
  \bibfield  {author} {\bibinfo {author} {\bibfnamefont {W.~N.}\ \bibnamefont
  {Faugno}}, \bibinfo {author} {\bibfnamefont {J.~K.}\ \bibnamefont {Jain}},\
  and\ \bibinfo {author} {\bibfnamefont {A.~C.}\ \bibnamefont {Balram}},\
  }\bibfield  {title} {\bibinfo {title} {Non-abelian fractional quantum {Hall}
  state at $3/7$-filled {Landau} level},\ }\href
  {https://doi.org/10.1103/PhysRevResearch.2.033223} {\bibfield  {journal}
  {\bibinfo  {journal} {Phys. Rev. Research}\ }\textbf {\bibinfo {volume}
  {2}},\ \bibinfo {pages} {033223} (\bibinfo {year} {2020})}\BibitemShut
  {NoStop}%
\bibitem [{\citenamefont {Balram}\ \emph {et~al.}(2020)\citenamefont {Balram},
  \citenamefont {Jain},\ and\ \citenamefont {Barkeshli}}]{Balram20}%
  \BibitemOpen
  \bibfield  {author} {\bibinfo {author} {\bibfnamefont {A.~C.}\ \bibnamefont
  {Balram}}, \bibinfo {author} {\bibfnamefont {J.~K.}\ \bibnamefont {Jain}},\
  and\ \bibinfo {author} {\bibfnamefont {M.}~\bibnamefont {Barkeshli}},\
  }\bibfield  {title} {\bibinfo {title} {${\mathbb{z}}_{n}$ superconductivity
  of composite bosons and the $7/3$ fractional quantum {Hall} effect},\ }\href
  {https://doi.org/10.1103/PhysRevResearch.2.013349} {\bibfield  {journal}
  {\bibinfo  {journal} {Phys. Rev. Research}\ }\textbf {\bibinfo {volume}
  {2}},\ \bibinfo {pages} {013349} (\bibinfo {year} {2020})}\BibitemShut
  {NoStop}%
\bibitem [{\citenamefont {Balram}\ and\ \citenamefont
  {W\'ojs}(2021)}]{Balram21a}%
  \BibitemOpen
  \bibfield  {author} {\bibinfo {author} {\bibfnamefont {A.~C.}\ \bibnamefont
  {Balram}}\ and\ \bibinfo {author} {\bibfnamefont {A.}~\bibnamefont
  {W\'ojs}},\ }\bibfield  {title} {\bibinfo {title} {Parton wave function for
  the fractional quantum hall effect at $\ensuremath{\nu}=6/17$},\ }\href
  {https://doi.org/10.1103/PhysRevResearch.3.033087} {\bibfield  {journal}
  {\bibinfo  {journal} {Phys. Rev. Research}\ }\textbf {\bibinfo {volume}
  {3}},\ \bibinfo {pages} {033087} (\bibinfo {year} {2021})}\BibitemShut
  {NoStop}%
\bibitem [{\citenamefont {Balram}(2022)}]{Balram21b}%
  \BibitemOpen
  \bibfield  {author} {\bibinfo {author} {\bibfnamefont {A.~C.}\ \bibnamefont
  {Balram}},\ }\bibfield  {title} {\bibinfo {title} {Transitions from {Abelian}
  composite fermion to non-{Abelian} parton fractional quantum {Hall} states in
  the zeroth {Landau} level of bilayer graphene},\ }\href
  {https://doi.org/10.1103/PhysRevB.105.L121406} {\bibfield  {journal}
  {\bibinfo  {journal} {Phys. Rev. B}\ }\textbf {\bibinfo {volume} {105}},\
  \bibinfo {pages} {L121406} (\bibinfo {year} {2022})}\BibitemShut {NoStop}%
\bibitem [{\citenamefont {Balram}(2021{\natexlab{b}})}]{Balram21c}%
  \BibitemOpen
  \bibfield  {author} {\bibinfo {author} {\bibfnamefont {A.~C.}\ \bibnamefont
  {Balram}},\ }\bibfield  {title} {\bibinfo {title} {Abelian parton state for
  the $\ensuremath{\nu}=4/11$ fractional quantum {Hall} effect},\ }\href
  {https://doi.org/10.1103/PhysRevB.103.155103} {\bibfield  {journal} {\bibinfo
   {journal} {Phys. Rev. B}\ }\textbf {\bibinfo {volume} {103}},\ \bibinfo
  {pages} {155103} (\bibinfo {year} {2021}{\natexlab{b}})}\BibitemShut
  {NoStop}%
\bibitem [{\citenamefont {Balram}\ \emph {et~al.}(2022)\citenamefont {Balram},
  \citenamefont {Liu}, \citenamefont {Gromov},\ and\ \citenamefont
  {Papi\ifmmode~\acute{c}\else \'{c}\fi{}}}]{Balram21d}%
  \BibitemOpen
  \bibfield  {author} {\bibinfo {author} {\bibfnamefont {A.~C.}\ \bibnamefont
  {Balram}}, \bibinfo {author} {\bibfnamefont {Z.}~\bibnamefont {Liu}},
  \bibinfo {author} {\bibfnamefont {A.}~\bibnamefont {Gromov}},\ and\ \bibinfo
  {author} {\bibfnamefont {Z.}~\bibnamefont {Papi\ifmmode~\acute{c}\else
  \'{c}\fi{}}},\ }\bibfield  {title} {\bibinfo {title} {Very-high-energy
  collective states of partons in fractional quantum {Hall} liquids},\ }\href
  {https://doi.org/10.1103/PhysRevX.12.021008} {\bibfield  {journal} {\bibinfo
  {journal} {Phys. Rev. X}\ }\textbf {\bibinfo {volume} {12}},\ \bibinfo
  {pages} {021008} (\bibinfo {year} {2022})}\BibitemShut {NoStop}%
\bibitem [{\citenamefont {Sharma}\ \emph {et~al.}(2023)\citenamefont {Sharma},
  \citenamefont {Pu}, \citenamefont {Balram},\ and\ \citenamefont
  {Jain}}]{Sharma22}%
  \BibitemOpen
  \bibfield  {author} {\bibinfo {author} {\bibfnamefont {A.}~\bibnamefont
  {Sharma}}, \bibinfo {author} {\bibfnamefont {S.}~\bibnamefont {Pu}}, \bibinfo
  {author} {\bibfnamefont {A.~C.}\ \bibnamefont {Balram}},\ and\ \bibinfo
  {author} {\bibfnamefont {J.~K.}\ \bibnamefont {Jain}},\ }\bibfield  {title}
  {\bibinfo {title} {Fractional quantum {Hall} effect with unconventional
  pairing in monolayer graphene},\ }\href
  {https://doi.org/10.1103/PhysRevLett.130.126201} {\bibfield  {journal}
  {\bibinfo  {journal} {Phys. Rev. Lett.}\ }\textbf {\bibinfo {volume} {130}},\
  \bibinfo {pages} {126201} (\bibinfo {year} {2023})}\BibitemShut {NoStop}%
\bibitem [{\citenamefont {Balram}\ and\ \citenamefont
  {Jain}(2016)}]{Balram16b}%
  \BibitemOpen
  \bibfield  {author} {\bibinfo {author} {\bibfnamefont {A.~C.}\ \bibnamefont
  {Balram}}\ and\ \bibinfo {author} {\bibfnamefont {J.~K.}\ \bibnamefont
  {Jain}},\ }\bibfield  {title} {\bibinfo {title} {Nature of composite fermions
  and the role of particle-hole symmetry: A microscopic account},\ }\href
  {https://doi.org/10.1103/PhysRevB.93.235152} {\bibfield  {journal} {\bibinfo
  {journal} {Phys. Rev. B}\ }\textbf {\bibinfo {volume} {93}},\ \bibinfo
  {pages} {235152} (\bibinfo {year} {2016})}\BibitemShut {NoStop}%
\bibitem [{\citenamefont {Anand}\ \emph {et~al.}(2022)\citenamefont {Anand},
  \citenamefont {Patil}, \citenamefont {Balram},\ and\ \citenamefont
  {Sreejith}}]{Anand22}%
  \BibitemOpen
  \bibfield  {author} {\bibinfo {author} {\bibfnamefont {A.}~\bibnamefont
  {Anand}}, \bibinfo {author} {\bibfnamefont {R.~A.}\ \bibnamefont {Patil}},
  \bibinfo {author} {\bibfnamefont {A.~C.}\ \bibnamefont {Balram}},\ and\
  \bibinfo {author} {\bibfnamefont {G.~J.}\ \bibnamefont {Sreejith}},\
  }\bibfield  {title} {\bibinfo {title} {Real-space entanglement spectra of
  parton states in fractional quantum {Hall} systems},\ }\href
  {https://doi.org/10.1103/PhysRevB.106.085136} {\bibfield  {journal} {\bibinfo
   {journal} {Phys. Rev. B}\ }\textbf {\bibinfo {volume} {106}},\ \bibinfo
  {pages} {085136} (\bibinfo {year} {2022})}\BibitemShut {NoStop}%
\bibitem [{\citenamefont {Balram}\ \emph
  {et~al.}(2018{\natexlab{b}})\citenamefont {Balram}, \citenamefont
  {Mukherjee}, \citenamefont {Park}, \citenamefont {Barkeshli}, \citenamefont
  {Rudner},\ and\ \citenamefont {Jain}}]{Balram18a}%
  \BibitemOpen
  \bibfield  {author} {\bibinfo {author} {\bibfnamefont {A.~C.}\ \bibnamefont
  {Balram}}, \bibinfo {author} {\bibfnamefont {S.}~\bibnamefont {Mukherjee}},
  \bibinfo {author} {\bibfnamefont {K.}~\bibnamefont {Park}}, \bibinfo {author}
  {\bibfnamefont {M.}~\bibnamefont {Barkeshli}}, \bibinfo {author}
  {\bibfnamefont {M.~S.}\ \bibnamefont {Rudner}},\ and\ \bibinfo {author}
  {\bibfnamefont {J.~K.}\ \bibnamefont {Jain}},\ }\bibfield  {title} {\bibinfo
  {title} {Fractional quantum {Hall} effect at $\ensuremath{\nu}=2+6/13$: The
  parton paradigm for the second {Landau} level},\ }\href
  {https://doi.org/10.1103/PhysRevLett.121.186601} {\bibfield  {journal}
  {\bibinfo  {journal} {Phys. Rev. Lett.}\ }\textbf {\bibinfo {volume} {121}},\
  \bibinfo {pages} {186601} (\bibinfo {year} {2018}{\natexlab{b}})}\BibitemShut
  {NoStop}%
\bibitem [{SM()}]{SM}%
  \BibitemOpen
  \href@noop {} {}\bibinfo {note} {See Supplemental Material that contains (i)
  the overlaps and gaps for the $\bar{3}\bar{2}1^{3}$ parton state at 6/13 in
  the zeroth Landau level of bilayer graphene and (ii) the overlaps and
  energy-variances of the quasiholes of the Moore-Read state, which includes
  Ref.~\cite{Greiter91}.}\BibitemShut {Stop}%
\bibitem [{\citenamefont {Balram}\ \emph
  {et~al.}(2015{\natexlab{b}})\citenamefont {Balram}, \citenamefont {T\"oke},
  \citenamefont {W\'ojs},\ and\ \citenamefont {Jain}}]{Balram15}%
  \BibitemOpen
  \bibfield  {author} {\bibinfo {author} {\bibfnamefont {A.~C.}\ \bibnamefont
  {Balram}}, \bibinfo {author} {\bibfnamefont {C.}~\bibnamefont {T\"oke}},
  \bibinfo {author} {\bibfnamefont {A.}~\bibnamefont {W\'ojs}},\ and\ \bibinfo
  {author} {\bibfnamefont {J.~K.}\ \bibnamefont {Jain}},\ }\bibfield  {title}
  {\bibinfo {title} {Phase diagram of fractional quantum {Hall} effect of
  composite fermions in multicomponent systems},\ }\href
  {https://doi.org/10.1103/PhysRevB.91.045109} {\bibfield  {journal} {\bibinfo
  {journal} {Phys. Rev. B}\ }\textbf {\bibinfo {volume} {91}},\ \bibinfo
  {pages} {045109} (\bibinfo {year} {2015}{\natexlab{b}})}\BibitemShut
  {NoStop}%
\bibitem [{\citenamefont {Pan}\ \emph {et~al.}(2003)\citenamefont {Pan},
  \citenamefont {Stormer}, \citenamefont {Tsui}, \citenamefont {Pfeiffer},
  \citenamefont {Baldwin},\ and\ \citenamefont {West}}]{Pan03}%
  \BibitemOpen
  \bibfield  {author} {\bibinfo {author} {\bibfnamefont {W.}~\bibnamefont
  {Pan}}, \bibinfo {author} {\bibfnamefont {H.~L.}\ \bibnamefont {Stormer}},
  \bibinfo {author} {\bibfnamefont {D.~C.}\ \bibnamefont {Tsui}}, \bibinfo
  {author} {\bibfnamefont {L.~N.}\ \bibnamefont {Pfeiffer}}, \bibinfo {author}
  {\bibfnamefont {K.~W.}\ \bibnamefont {Baldwin}},\ and\ \bibinfo {author}
  {\bibfnamefont {K.~W.}\ \bibnamefont {West}},\ }\bibfield  {title} {\bibinfo
  {title} {Fractional quantum {Hall} effect of composite fermions},\ }\href
  {https://doi.org/10.1103/PhysRevLett.90.016801} {\bibfield  {journal}
  {\bibinfo  {journal} {Phys. Rev. Lett.}\ }\textbf {\bibinfo {volume} {90}},\
  \bibinfo {pages} {016801} (\bibinfo {year} {2003})}\BibitemShut {NoStop}%
\bibitem [{\citenamefont {Samkharadze}\ \emph {et~al.}(2015)\citenamefont
  {Samkharadze}, \citenamefont {Arnold}, \citenamefont {Pfeiffer},
  \citenamefont {West},\ and\ \citenamefont {Cs\'athy}}]{Samkharadze15b}%
  \BibitemOpen
  \bibfield  {author} {\bibinfo {author} {\bibfnamefont {N.}~\bibnamefont
  {Samkharadze}}, \bibinfo {author} {\bibfnamefont {I.}~\bibnamefont {Arnold}},
  \bibinfo {author} {\bibfnamefont {L.~N.}\ \bibnamefont {Pfeiffer}}, \bibinfo
  {author} {\bibfnamefont {K.~W.}\ \bibnamefont {West}},\ and\ \bibinfo
  {author} {\bibfnamefont {G.~A.}\ \bibnamefont {Cs\'athy}},\ }\bibfield
  {title} {\bibinfo {title} {Observation of incompressibility at $\nu=4/11$ and
  $\nu=5/13$},\ }\href {https://doi.org/10.1103/PhysRevB.91.081109} {\bibfield
  {journal} {\bibinfo  {journal} {Phys. Rev. B}\ }\textbf {\bibinfo {volume}
  {91}},\ \bibinfo {pages} {081109} (\bibinfo {year} {2015})}\BibitemShut
  {NoStop}%
\bibitem [{\citenamefont {Pan}\ \emph {et~al.}(2015)\citenamefont {Pan},
  \citenamefont {Baldwin}, \citenamefont {West}, \citenamefont {Pfeiffer},\
  and\ \citenamefont {Tsui}}]{Pan15}%
  \BibitemOpen
  \bibfield  {author} {\bibinfo {author} {\bibfnamefont {W.}~\bibnamefont
  {Pan}}, \bibinfo {author} {\bibfnamefont {K.~W.}\ \bibnamefont {Baldwin}},
  \bibinfo {author} {\bibfnamefont {K.~W.}\ \bibnamefont {West}}, \bibinfo
  {author} {\bibfnamefont {L.~N.}\ \bibnamefont {Pfeiffer}},\ and\ \bibinfo
  {author} {\bibfnamefont {D.~C.}\ \bibnamefont {Tsui}},\ }\bibfield  {title}
  {\bibinfo {title} {Fractional quantum {Hall} effect at {Landau} level filling
  $\ensuremath{\nu}=4/11$},\ }\href
  {https://doi.org/10.1103/PhysRevB.91.041301} {\bibfield  {journal} {\bibinfo
  {journal} {Phys. Rev. B}\ }\textbf {\bibinfo {volume} {91}},\ \bibinfo
  {pages} {041301} (\bibinfo {year} {2015})}\BibitemShut {NoStop}%
\bibitem [{\citenamefont {Kumar}\ \emph {et~al.}(2019)\citenamefont {Kumar},
  \citenamefont {Raghu},\ and\ \citenamefont {Mulligan}}]{Kumar18}%
  \BibitemOpen
  \bibfield  {author} {\bibinfo {author} {\bibfnamefont {P.}~\bibnamefont
  {Kumar}}, \bibinfo {author} {\bibfnamefont {S.}~\bibnamefont {Raghu}},\ and\
  \bibinfo {author} {\bibfnamefont {M.}~\bibnamefont {Mulligan}},\ }\bibfield
  {title} {\bibinfo {title} {Composite fermion {Hall} conductivity and the
  half-filled {Landau} level},\ }\href
  {https://doi.org/10.1103/PhysRevB.99.235114} {\bibfield  {journal} {\bibinfo
  {journal} {Phys. Rev. B}\ }\textbf {\bibinfo {volume} {99}},\ \bibinfo
  {pages} {235114} (\bibinfo {year} {2019})}\BibitemShut {NoStop}%
\bibitem [{\citenamefont {Dora}\ and\ \citenamefont {Balram}(2022)}]{Dora22}%
  \BibitemOpen
  \bibfield  {author} {\bibinfo {author} {\bibfnamefont {R.~K.}\ \bibnamefont
  {Dora}}\ and\ \bibinfo {author} {\bibfnamefont {A.~C.}\ \bibnamefont
  {Balram}},\ }\bibfield  {title} {\bibinfo {title} {Nature of the anomalous
  $4/13$ fractional quantum {Hall} effect in graphene},\ }\href
  {https://doi.org/10.1103/PhysRevB.105.L241403} {\bibfield  {journal}
  {\bibinfo  {journal} {Phys. Rev. B}\ }\textbf {\bibinfo {volume} {105}},\
  \bibinfo {pages} {L241403} (\bibinfo {year} {2022})}\BibitemShut {NoStop}%
\bibitem [{\citenamefont {Jung}\ and\ \citenamefont
  {MacDonald}(2014)}]{Jung14}%
  \BibitemOpen
  \bibfield  {author} {\bibinfo {author} {\bibfnamefont {J.}~\bibnamefont
  {Jung}}\ and\ \bibinfo {author} {\bibfnamefont {A.~H.}\ \bibnamefont
  {MacDonald}},\ }\bibfield  {title} {\bibinfo {title} {Accurate tight-binding
  models for the $\ensuremath{\pi}$ bands of bilayer graphene},\ }\href
  {https://doi.org/10.1103/PhysRevB.89.035405} {\bibfield  {journal} {\bibinfo
  {journal} {Phys. Rev. B}\ }\textbf {\bibinfo {volume} {89}},\ \bibinfo
  {pages} {035405} (\bibinfo {year} {2014})}\BibitemShut {NoStop}%
\bibitem [{\citenamefont {Faugno}\ \emph {et~al.}(2021)\citenamefont {Faugno},
  \citenamefont {Zhao}, \citenamefont {Balram}, \citenamefont {Jolicoeur},\
  and\ \citenamefont {Jain}}]{Faugno21}%
  \BibitemOpen
  \bibfield  {author} {\bibinfo {author} {\bibfnamefont {W.~N.}\ \bibnamefont
  {Faugno}}, \bibinfo {author} {\bibfnamefont {T.}~\bibnamefont {Zhao}},
  \bibinfo {author} {\bibfnamefont {A.~C.}\ \bibnamefont {Balram}}, \bibinfo
  {author} {\bibfnamefont {T.}~\bibnamefont {Jolicoeur}},\ and\ \bibinfo
  {author} {\bibfnamefont {J.~K.}\ \bibnamefont {Jain}},\ }\bibfield  {title}
  {\bibinfo {title} {Unconventional ${\mathbb{z}}_{n}$ parton states at
  $\ensuremath{\nu}=7/3$: {Role} of finite width},\ }\href
  {https://doi.org/10.1103/PhysRevB.103.085303} {\bibfield  {journal} {\bibinfo
   {journal} {Phys. Rev. B}\ }\textbf {\bibinfo {volume} {103}},\ \bibinfo
  {pages} {085303} (\bibinfo {year} {2021})}\BibitemShut {NoStop}%
\bibitem [{\citenamefont {Sreejith}\ \emph {et~al.}(2013)\citenamefont
  {Sreejith}, \citenamefont {Wu}, \citenamefont {W\'ojs},\ and\ \citenamefont
  {Jain}}]{Sreejith13}%
  \BibitemOpen
  \bibfield  {author} {\bibinfo {author} {\bibfnamefont {G.~J.}\ \bibnamefont
  {Sreejith}}, \bibinfo {author} {\bibfnamefont {Y.-H.}\ \bibnamefont {Wu}},
  \bibinfo {author} {\bibfnamefont {A.}~\bibnamefont {W\'ojs}},\ and\ \bibinfo
  {author} {\bibfnamefont {J.~K.}\ \bibnamefont {Jain}},\ }\bibfield  {title}
  {\bibinfo {title} {Tripartite composite fermion states},\ }\href
  {https://doi.org/10.1103/PhysRevB.87.245125} {\bibfield  {journal} {\bibinfo
  {journal} {Phys. Rev. B}\ }\textbf {\bibinfo {volume} {87}},\ \bibinfo
  {pages} {245125} (\bibinfo {year} {2013})}\BibitemShut {NoStop}%
\bibitem [{\citenamefont {Zibrov}\ \emph {et~al.}(2018)\citenamefont {Zibrov},
  \citenamefont {Spanton}, \citenamefont {Zhou}, \citenamefont {Kometter},
  \citenamefont {Taniguchi}, \citenamefont {Watanabe},\ and\ \citenamefont
  {Young}}]{Zibrov17}%
  \BibitemOpen
  \bibfield  {author} {\bibinfo {author} {\bibfnamefont {A.~A.}\ \bibnamefont
  {Zibrov}}, \bibinfo {author} {\bibfnamefont {E.~M.}\ \bibnamefont {Spanton}},
  \bibinfo {author} {\bibfnamefont {H.}~\bibnamefont {Zhou}}, \bibinfo {author}
  {\bibfnamefont {C.}~\bibnamefont {Kometter}}, \bibinfo {author}
  {\bibfnamefont {T.}~\bibnamefont {Taniguchi}}, \bibinfo {author}
  {\bibfnamefont {K.}~\bibnamefont {Watanabe}},\ and\ \bibinfo {author}
  {\bibfnamefont {A.~F.}\ \bibnamefont {Young}},\ }\bibfield  {title} {\bibinfo
  {title} {Even-denominator fractional quantum {Hall} states at an isospin
  transition in monolayer graphene},\ }\href
  {https://doi.org/10.1038/s41567-018-0190-0} {\bibfield  {journal} {\bibinfo
  {journal} {Nature Physics}\ }\textbf {\bibinfo {volume} {14}},\ \bibinfo
  {pages} {930} (\bibinfo {year} {2018})}\BibitemShut {NoStop}%
\bibitem [{\citenamefont {d'Ambrumenil}\ and\ \citenamefont
  {Morf}(1989)}]{Ambrumenil89}%
  \BibitemOpen
  \bibfield  {author} {\bibinfo {author} {\bibfnamefont {N.}~\bibnamefont
  {d'Ambrumenil}}\ and\ \bibinfo {author} {\bibfnamefont {R.}~\bibnamefont
  {Morf}},\ }\bibfield  {title} {\bibinfo {title} {Hierarchical classification
  of fractional quantum {Hall} states},\ }\href
  {https://doi.org/10.1103/PhysRevB.40.6108} {\bibfield  {journal} {\bibinfo
  {journal} {Phys. Rev. B}\ }\textbf {\bibinfo {volume} {40}},\ \bibinfo
  {pages} {6108} (\bibinfo {year} {1989})}\BibitemShut {NoStop}%
\bibitem [{\citenamefont {Wen}(1991)}]{Wen91b}%
  \BibitemOpen
  \bibfield  {author} {\bibinfo {author} {\bibfnamefont {X.}~\bibnamefont
  {Wen}},\ }\bibfield  {title} {\bibinfo {title} {Edge excitations in the
  fractional quantum {Hall} states at general filling fractions},\ }\href
  {https://doi.org/10.1142/S0217984991000058} {\bibfield  {journal} {\bibinfo
  {journal} {Modern Physics Letters B}\ }\textbf {\bibinfo {volume} {05}},\
  \bibinfo {pages} {39} (\bibinfo {year} {1991})}\BibitemShut {NoStop}%
\bibitem [{\citenamefont {Wen}(1992)}]{Wen92b}%
  \BibitemOpen
  \bibfield  {author} {\bibinfo {author} {\bibfnamefont {X.-G.}\ \bibnamefont
  {Wen}},\ }\bibfield  {title} {\bibinfo {title} {Theory of the edge states in
  fractional quantum {Hall} effects},\ }\href
  {https://doi.org/10.1142/S0217979292000840} {\bibfield  {journal} {\bibinfo
  {journal} {International Journal of Modern Physics B}\ }\textbf {\bibinfo
  {volume} {06}},\ \bibinfo {pages} {1711} (\bibinfo {year}
  {1992})}\BibitemShut {NoStop}%
\bibitem [{\citenamefont {Moore}\ and\ \citenamefont {Wen}(1998)}]{Moore98}%
  \BibitemOpen
  \bibfield  {author} {\bibinfo {author} {\bibfnamefont {J.~E.}\ \bibnamefont
  {Moore}}\ and\ \bibinfo {author} {\bibfnamefont {X.-G.}\ \bibnamefont
  {Wen}},\ }\bibfield  {title} {\bibinfo {title} {Classification of disordered
  phases of quantum {Hall} edge states},\ }\href
  {https://doi.org/10.1103/PhysRevB.57.10138} {\bibfield  {journal} {\bibinfo
  {journal} {Phys. Rev. B}\ }\textbf {\bibinfo {volume} {57}},\ \bibinfo
  {pages} {10138} (\bibinfo {year} {1998})}\BibitemShut {NoStop}%
\bibitem [{\citenamefont {Bid}\ \emph {et~al.}(2010)\citenamefont {Bid},
  \citenamefont {Ofek}, \citenamefont {Inoue}, \citenamefont {Heiblum},
  \citenamefont {Kane}, \citenamefont {Umansky},\ and\ \citenamefont
  {Mahalu}}]{Bid10}%
  \BibitemOpen
  \bibfield  {author} {\bibinfo {author} {\bibfnamefont {A.}~\bibnamefont
  {Bid}}, \bibinfo {author} {\bibfnamefont {N.}~\bibnamefont {Ofek}}, \bibinfo
  {author} {\bibfnamefont {H.}~\bibnamefont {Inoue}}, \bibinfo {author}
  {\bibfnamefont {M.}~\bibnamefont {Heiblum}}, \bibinfo {author} {\bibfnamefont
  {C.~L.}\ \bibnamefont {Kane}}, \bibinfo {author} {\bibfnamefont
  {V.}~\bibnamefont {Umansky}},\ and\ \bibinfo {author} {\bibfnamefont
  {D.}~\bibnamefont {Mahalu}},\ }\bibfield  {title} {\bibinfo {title}
  {Observation of neutral modes in the fractional quantum {Hall} regime},\
  }\href {https://doi.org/10.1038/nature09277} {\bibfield  {journal} {\bibinfo
  {journal} {Nature}\ }\textbf {\bibinfo {volume} {466}},\ \bibinfo {pages}
  {585} (\bibinfo {year} {2010})}\BibitemShut {NoStop}%
\bibitem [{\citenamefont {Dolev}\ \emph {et~al.}(2011)\citenamefont {Dolev},
  \citenamefont {Gross}, \citenamefont {Sabo}, \citenamefont {Gurman},
  \citenamefont {Heiblum}, \citenamefont {Umansky},\ and\ \citenamefont
  {Mahalu}}]{Dolev11}%
  \BibitemOpen
  \bibfield  {author} {\bibinfo {author} {\bibfnamefont {M.}~\bibnamefont
  {Dolev}}, \bibinfo {author} {\bibfnamefont {Y.}~\bibnamefont {Gross}},
  \bibinfo {author} {\bibfnamefont {R.}~\bibnamefont {Sabo}}, \bibinfo {author}
  {\bibfnamefont {I.}~\bibnamefont {Gurman}}, \bibinfo {author} {\bibfnamefont
  {M.}~\bibnamefont {Heiblum}}, \bibinfo {author} {\bibfnamefont
  {V.}~\bibnamefont {Umansky}},\ and\ \bibinfo {author} {\bibfnamefont
  {D.}~\bibnamefont {Mahalu}},\ }\bibfield  {title} {\bibinfo {title}
  {Characterizing neutral modes of fractional states in the second {Landau}
  level},\ }\href {https://doi.org/10.1103/PhysRevLett.107.036805} {\bibfield
  {journal} {\bibinfo  {journal} {Phys. Rev. Lett.}\ }\textbf {\bibinfo
  {volume} {107}},\ \bibinfo {pages} {036805} (\bibinfo {year}
  {2011})}\BibitemShut {NoStop}%
\bibitem [{\citenamefont {Banerjee}\ \emph {et~al.}(2017)\citenamefont
  {Banerjee}, \citenamefont {Heiblum}, \citenamefont {Rosenblatt},
  \citenamefont {Oreg}, \citenamefont {Feldman}, \citenamefont {Stern},\ and\
  \citenamefont {Umansky}}]{Banerjee17}%
  \BibitemOpen
  \bibfield  {author} {\bibinfo {author} {\bibfnamefont {M.}~\bibnamefont
  {Banerjee}}, \bibinfo {author} {\bibfnamefont {M.}~\bibnamefont {Heiblum}},
  \bibinfo {author} {\bibfnamefont {A.}~\bibnamefont {Rosenblatt}}, \bibinfo
  {author} {\bibfnamefont {Y.}~\bibnamefont {Oreg}}, \bibinfo {author}
  {\bibfnamefont {D.~E.}\ \bibnamefont {Feldman}}, \bibinfo {author}
  {\bibfnamefont {A.}~\bibnamefont {Stern}},\ and\ \bibinfo {author}
  {\bibfnamefont {V.}~\bibnamefont {Umansky}},\ }\bibfield  {title} {\bibinfo
  {title} {Observed quantization of anyonic heat flow},\ }\href
  {https://doi.org/10.1038/nature22052} {\bibfield  {journal} {\bibinfo
  {journal} {Nature}\ }\textbf {\bibinfo {volume} {545}},\ \bibinfo {pages}
  {75} (\bibinfo {year} {2017})}\BibitemShut {NoStop}%
\bibitem [{\citenamefont {Banerjee}\ \emph {et~al.}(2018)\citenamefont
  {Banerjee}, \citenamefont {Heiblum}, \citenamefont {Umansky}, \citenamefont
  {Feldman}, \citenamefont {Oreg},\ and\ \citenamefont {Stern}}]{Banerjee18}%
  \BibitemOpen
  \bibfield  {author} {\bibinfo {author} {\bibfnamefont {M.}~\bibnamefont
  {Banerjee}}, \bibinfo {author} {\bibfnamefont {M.}~\bibnamefont {Heiblum}},
  \bibinfo {author} {\bibfnamefont {V.}~\bibnamefont {Umansky}}, \bibinfo
  {author} {\bibfnamefont {D.~E.}\ \bibnamefont {Feldman}}, \bibinfo {author}
  {\bibfnamefont {Y.}~\bibnamefont {Oreg}},\ and\ \bibinfo {author}
  {\bibfnamefont {A.}~\bibnamefont {Stern}},\ }\bibfield  {title} {\bibinfo
  {title} {Observation of half-integer thermal hall conductance},\ }\href@noop
  {} {\bibfield  {journal} {\bibinfo  {journal} {Nature}\ }\textbf {\bibinfo
  {volume} {559}},\ \bibinfo {pages} {205} (\bibinfo {year}
  {2018})}\BibitemShut {NoStop}%
\bibitem [{\citenamefont {Srivastav}\ \emph {et~al.}(2019)\citenamefont
  {Srivastav}, \citenamefont {Sahu}, \citenamefont {Watanabe}, \citenamefont
  {Taniguchi}, \citenamefont {Banerjee},\ and\ \citenamefont
  {Das}}]{Srivastav19}%
  \BibitemOpen
  \bibfield  {author} {\bibinfo {author} {\bibfnamefont {S.~K.}\ \bibnamefont
  {Srivastav}}, \bibinfo {author} {\bibfnamefont {M.~R.}\ \bibnamefont {Sahu}},
  \bibinfo {author} {\bibfnamefont {K.}~\bibnamefont {Watanabe}}, \bibinfo
  {author} {\bibfnamefont {T.}~\bibnamefont {Taniguchi}}, \bibinfo {author}
  {\bibfnamefont {S.}~\bibnamefont {Banerjee}},\ and\ \bibinfo {author}
  {\bibfnamefont {A.}~\bibnamefont {Das}},\ }\bibfield  {title} {\bibinfo
  {title} {Universal quantized thermal conductance in graphene},\ }\bibfield
  {journal} {\bibinfo  {journal} {Science Advances}\ }\textbf {\bibinfo
  {volume} {5}},\ \href {https://doi.org/10.1126/sciadv.aaw5798}
  {10.1126/sciadv.aaw5798} (\bibinfo {year} {2019})\BibitemShut {NoStop}%
\bibitem [{\citenamefont {Read}(2009)}]{Read09}%
  \BibitemOpen
  \bibfield  {author} {\bibinfo {author} {\bibfnamefont {N.}~\bibnamefont
  {Read}},\ }\bibfield  {title} {\bibinfo {title} {Non-abelian adiabatic
  statistics and {Hall} viscosity in quantum {Hall} states and
  ${p}_{x}+i{p}_{y}$ paired superfluids},\ }\href
  {https://doi.org/10.1103/PhysRevB.79.045308} {\bibfield  {journal} {\bibinfo
  {journal} {Phys. Rev. B}\ }\textbf {\bibinfo {volume} {79}},\ \bibinfo
  {pages} {045308} (\bibinfo {year} {2009})}\BibitemShut {NoStop}%
\bibitem [{\citenamefont {Yutushui}\ \emph {et~al.}(2024)\citenamefont
  {Yutushui}, \citenamefont {Hermanns},\ and\ \citenamefont
  {Mross}}]{Yutushui24}%
  \BibitemOpen
  \bibfield  {author} {\bibinfo {author} {\bibfnamefont {M.}~\bibnamefont
  {Yutushui}}, \bibinfo {author} {\bibfnamefont {M.}~\bibnamefont {Hermanns}},\
  and\ \bibinfo {author} {\bibfnamefont {D.~F.}\ \bibnamefont {Mross}},\
  }\href@noop {} {\bibinfo {title} {Paired fermions in strong magnetic fields
  and daughters of even-denominator {Hall} plateaus}} (\bibinfo {year}
  {2024}),\ \Eprint {https://arxiv.org/abs/2405.03753} {arXiv:2405.03753
  [cond-mat.str-el]} \BibitemShut {NoStop}%
\bibitem [{\citenamefont {Zheltonozhskii}\ \emph {et~al.}(2024)\citenamefont
  {Zheltonozhskii}, \citenamefont {Stern},\ and\ \citenamefont
  {Lindner}}]{Zheltonozhskii24}%
  \BibitemOpen
  \bibfield  {author} {\bibinfo {author} {\bibfnamefont {E.}~\bibnamefont
  {Zheltonozhskii}}, \bibinfo {author} {\bibfnamefont {A.}~\bibnamefont
  {Stern}},\ and\ \bibinfo {author} {\bibfnamefont {N.}~\bibnamefont
  {Lindner}},\ }\href@noop {} {\bibinfo {title} {Identifying the topological
  order of quantized half-filled {Landau} levels through their daughter
  states}} (\bibinfo {year} {2024}),\ \Eprint
  {https://arxiv.org/abs/2405.03780} {arXiv:2405.03780 [cond-mat.mes-hall]}
  \BibitemShut {NoStop}%
\bibitem [{dia()}]{diagham}%
  \BibitemOpen
  \href@noop {} {}\bibinfo {note} {Diag{H}am,
  \url{https://www.nick-ux.org/diagham}}\BibitemShut {NoStop}%
\bibitem [{\citenamefont {Greiter}\ \emph {et~al.}(1991)\citenamefont
  {Greiter}, \citenamefont {Wen},\ and\ \citenamefont {Wilczek}}]{Greiter91}%
  \BibitemOpen
  \bibfield  {author} {\bibinfo {author} {\bibfnamefont {M.}~\bibnamefont
  {Greiter}}, \bibinfo {author} {\bibfnamefont {X.-G.}\ \bibnamefont {Wen}},\
  and\ \bibinfo {author} {\bibfnamefont {F.}~\bibnamefont {Wilczek}},\
  }\bibfield  {title} {\bibinfo {title} {Paired {Hall} state at half filling},\
  }\href {https://doi.org/10.1103/PhysRevLett.66.3205} {\bibfield  {journal}
  {\bibinfo  {journal} {Phys. Rev. Lett.}\ }\textbf {\bibinfo {volume} {66}},\
  \bibinfo {pages} {3205} (\bibinfo {year} {1991})}\BibitemShut {NoStop}%
\end{thebibliography}%

\newpage 
\cleardoublepage

\begin{center}
\textbf{\large Supplemental Material for ``Fractional quantum Hall effect of partons and the nature of the 8/17 state in the zeroth Landau level of bilayer graphene" }\\[5pt]
\vspace{0.1cm}
\begin{quote}
{\small In this Supplemental Material, we present (i) the overlaps and gaps for the $\bar{3}\bar{2}1^{3}$ parton state at 6/13 in the zeroth Landau level of bilayer graphene and (ii) the overlaps and energy-variances of the quasiholes of the Moore-Read state.
}\\[20pt]
\end{quote}
\end{center}

\setcounter{equation}{0}
\setcounter{figure}{0}
\setcounter{table}{0}
\setcounter{page}{1}
\setcounter{section}{0}
\makeatletter
\renewcommand{\theequation}{S\arabic{equation}}
\renewcommand{\thefigure}{S\arabic{figure}}
\renewcommand{\thesection}{S\Roman{section}}
\renewcommand{\thepage}{\arabic{page}}
\renewcommand{\thetable}{S\arabic{table}}

\vspace{0cm}

\section{$\bar{3}\bar{2}1^{3}$ state in the zeroth Landau level of bilayer graphene at $\nu{=}6/13$}

\begin{figure}[htpb]
	\begin{center}
		\includegraphics[width=0.49\textwidth,height=0.24\textwidth]{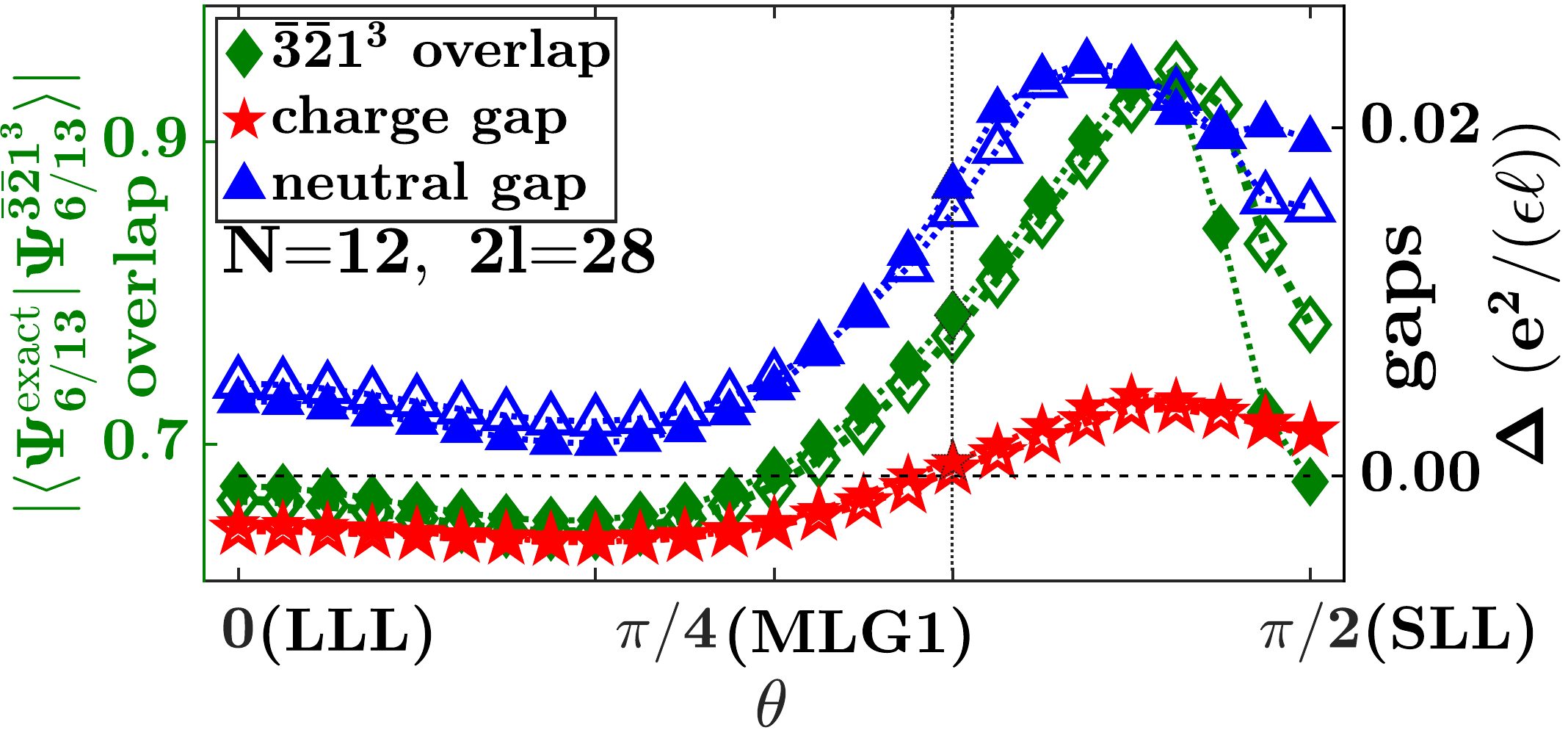} 
		\caption{(color online) Overlaps and gaps of the $\bar{3}\bar{2}1^{3}$ parton state with the exact Coulomb ground state in the zeroth Landau level of bilayer graphene at $\nu{=}6/13$ evaluated in the spherical geometry using the spherical (filled symbols) and disk (open symbols) pseudopotentials for $N{=}12$ electrons at flux $2l{=}28$ as a function of the parameter $\theta$, which is related to the perpendicular magnetic field $B{=}93.06[\cot(\theta)]^{2}$ (see text). The charge gaps are positive only for $\theta{\geq}\pi/3$, which indicates that the $\bar{3}\bar{2}1^{3}$ state is likely to be stabilized only in this region (to the right of the vertical dashed line).}
		\label{fig: overlaps_gaps_6_13_bar3bar2111_ZLL_BLG}
	\end{center}
\end{figure}

In Fig.~\ref{fig: overlaps_gaps_6_13_bar3bar2111_ZLL_BLG} we show overlaps of the $\bar{3}\bar{2}1^{3}$ with the exact Coulomb ground state in the ZLL of BLG. These overlaps are sizable for all values of $\theta$ in the ZLL of BLG. In Fig.~\ref{fig: overlaps_gaps_6_13_bar3bar2111_ZLL_BLG} we have also shown the charge and neutral gaps of the $\bar{3}\bar{2}1^{3}$ in the ZLL of BLG. The charge gap is defined as the energy to create a pair of fundamental quasiparticle-quasihole (the smallest charged quasihole in the $\bar{3}\bar{2}1^{3}$ has a charge $e/13$~\cite{Balram18a}). The neutral gap is defined as the energy difference between the ground state and the lowest-lying excitation. Both the gaps are positive only for $\theta{\geq}\pi/3$ which suggests that the $\bar{3}\bar{2}1^{3}$ state is stabilized only in this region. We note that in the thermodynamic limit, we expect the charge gap ${\geq}$ neutral gap but due to strong finite-size effects (routinely seen in the vicinity of the SLL) the ordering is opposite in Fig.~\ref{fig: overlaps_gaps_6_13_bar3bar2111_ZLL_BLG}.

\section{quasiholes of the Moore-Read state}
The Levin-Halperin state at 8/17 is obtained when the Moore-Read quasiholes have strong repulsion between them and they go on to condense to form a Laughlin state. Although we can construct the quasiholes of the Moore-Read state, it is not clear how to obtain the Levin-Halperin state from it. Indeed, the interactions that we diagonalize are between the parent electrons. Numerical construction of the LH state would require an interaction that is hard-core for the Moore-Read quasiholes whose expression in the parent electron language is still unknown.

\begin{figure}[htpb]
	\begin{center}
		\includegraphics[width=0.49\textwidth,height=0.27\textwidth]{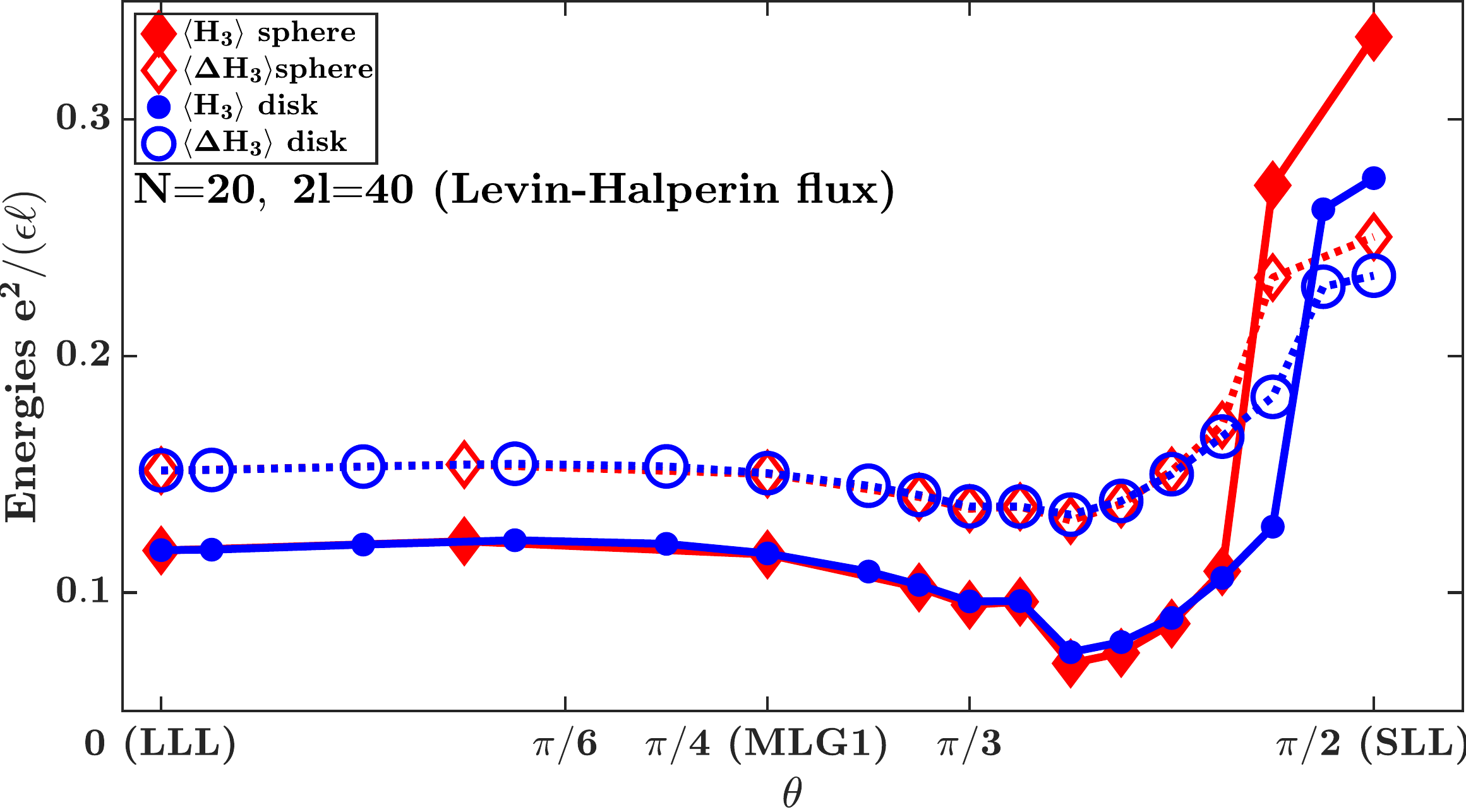} 
		\caption{(color online) The mean (filled symbols) and standard deviation (open symbols) of the exact Coulomb ground state in the zeroth Landau level of bilayer graphene for the three-body hard-core Hamiltonian $H_{3}$ at $\nu{=}8/17$ evaluated in the spherical geometry using the spherical (red diamonds) and disk (blue circles) pseudopotentials for $N{=}20$ electrons at flux $2l{=}40$ (Levin-Halperin flux) as a function of the parameter $\theta$.}
		\label{fig: H3_mean_std_dev_ZLL_BLG_N_20}
	\end{center}
\end{figure}

Nevertheless, the exact LH state should lie within the Moore-Read quasihole manifold. Thus, we have computed the overlaps of the Moore-Read quasiholes with the exact Coulomb ground state in the zeroth LL of bilayer graphene. Such a calculation can only be performed at $N{=}12$. As mentioned in the main text, the ground state in the ZLL of BLG has $L{=}0$ for all $\theta$. There are exactly 3 Moore-Read quasiholes states with $L{=}0$ at $N{=}12$ and $2l{=}23$ flux quanta. While the quasihole overlap calculation is out of reach for $N{=}20$, we have also computed as a proxy the $\langle H_{3} \rangle$ and $\langle \Delta H_{3} \rangle$, where $H_{3}$ is the model 3-body interaction which realizes the Moore-Read Pfaffian state~\cite{Greiter91}, of the exact Coulomb ground state in the zeroth LL of bilayer graphene for $N{=}20$. For the exact LH state, we would get $\langle H_{3} \rangle{=}\langle \Delta H_{3} \rangle{=}0$. These results are shown in Fig.~\ref{fig: H3_mean_std_dev_ZLL_BLG_N_20}. We find that both the $H_{3}$-mean and its standard deviation for the exact Coulomb ground state exhibit a minimum in the region where the ground state is uniform. For completeness, we point out that such calculations are demanding. To obtain the Coulomb ground state of the bilayer graphene for a given $\theta$, the average clock running time is 130h on two nodes with dual AMD EPYC 7542 processors (32 cores per CPU) with 2Tb of RAM each. On the other hand, an expectation value of $H_{3}$ requires on the order 250h using one such node.

Our results show that the Moore-Read quasiholes give a good description of the exact Coulomb ground state in the same region where the exact Coulomb ground state of $N{=}20$ at the Levin-Halperin flux at 8/17 has a uniform ground state, i.e., has $\langle L \rangle{=}0$ [see Fig. 2(b) of the main text]. However, this is also the region where the 1/2 Moore-Read state gives a good description of the exact Coulomb ground state in the half-filled zeroth LL of bilayer graphene~\cite{Balram21b}. For the Levin-Halperin state to be realized, the Moore-Read quasiholes need to have a strong repulsion between them and condense into a Laughlin state. Unfortunately, we do not have strong evidence to suggest that happens except for the fact that the full electronic ground state for $N{=}20$ has $\langle L \rangle{=}0$ in the same region where the overlap with the Moore-Read quasiholes is high.

\end{document}